\DeclareUnicodeCharacter{2061}{}
\documentclass[letter]{cas-sc}

\usepackage{placeins}
\usepackage{graphicx}

\usepackage[authoryear]{natbib}
\usepackage{booktabs} 
\usepackage{amsmath}
\usepackage{bbding}
\usepackage{amsmath}
\usepackage{booktabs} 
\usepackage{algorithm}
\usepackage{algpseudocode}
\usepackage[utf8]{inputenc}
\usepackage {booktabs}
\usepackage{enumitem}
\usepackage{orcidlink}
\usepackage{svg}
\usepackage{tabularx}
\usepackage{ragged2e}
\usepackage{subcaption}

\usepackage{array}
\newcolumntype{P}[1]{>{\centering\arraybackslash}p{#1}}
\newcolumntype{M}[1]{>{\centering\arraybackslash}m{#1}}
\usepackage{graphicx}
\usepackage{booktabs}  
\usepackage{multirow}   
\usepackage{ragged2e}
\usepackage{tikz}
\usetikzlibrary{shapes.geometric, arrows, positioning}

\usepackage{caption}

\hypersetup{
    colorlinks = true,
            linkcolor = blue,
            urlcolor  = blue,
            citecolor = blue,
            anchorcolor = blue,
    }
\def\tsc#1{\csdef{#1}{\textsc{\lowercase{#1}}\xspace}}
\tsc{WGM}
\tsc{QE}
\tsc{EP}
\tsc{PMS}
\tsc{BEC}
\tsc{DE}

\newcommand{\eat}[1]{}

\definecolor{color1}{RGB}{255, 204, 204}   
\definecolor{color2}{RGB}{255, 229, 204}   
\definecolor{color3}{RGB}{255, 255, 204}   
\definecolor{color4}{RGB}{229, 255, 204}   
\definecolor{color5}{RGB}{204, 255, 204}   
\definecolor{color6}{RGB}{204, 255, 229}   
\definecolor{color7}{RGB}{204, 255, 255}   
\definecolor{color8}{RGB}{204, 229, 255}   
\definecolor{color9}{RGB}{204, 204, 255}   
\definecolor{color10}{RGB}{229, 204, 255}  
\definecolor{color11}{RGB}{255, 204, 255}  
\definecolor{color12}{RGB}{255, 204, 229}  
\definecolor{color13}{RGB}{255, 102, 102}  
\definecolor{color14}{RGB}{255, 178, 102}  
\definecolor{color15}{RGB}{204, 204, 0}    
\definecolor{color16}{RGB}{102, 153, 0}    
\definecolor{color17}{RGB}{0, 153, 153}    
\definecolor{color18}{RGB}{0, 102, 204}    
\definecolor{color19}{RGB}{102, 0, 204}    
\definecolor{color20}{RGB}{153, 0, 76}     

\makeatletter
\floatstyle{plaintop}
\restylefloat{table}
\makeatother

\begin{document}
\let\WriteBookmarks\relax
\def\floatpagepagefraction{1}
\def\textpagefraction{.001}

\shorttitle{KnoVo}

\shortauthors{Rubaiat, Sakib and Jamil}

\title [mode = title]{
Mapping the Evolution of Research Contributions using KnoVo
}                      

%


\affiliation[1]{organization={
Department of Computer Science, University of Idaho},
    city={Mowcow},
    state={Idaho},
    postcode={83844},
    country={USA}}

\author[1]{Sajratul Y. Rubaiat}[
 style=chinese, orcid=0009-0001-0367-644X]

\author[1]{Syed N. Sakib}[
 style=chinese, orcid=0009-0000-2653-2767]

\author[1]{Hasan M. Jamil}[
 style=chinese, orcid=0000-0002-3124-3780]

\cormark[1]
\ead{jamil@uidaho.edu}



\cortext[cor1]{Corresponding author}

\begin{abstract}
This paper presents \textsc{KnoVo} (Knowledge Evolution), an intelligent framework designed for \emph{quantifying} and analyzing the evolution of research novelty in the scientific literature. Moving beyond traditional citation analysis, which primarily measures impact, KnoVo determines a paper's novelty relative to both prior and subsequent work within its multilayered citation network. Given a target paper's abstract, KnoVo utilizes Large Language Models (LLMs) to dynamically extract \emph{dimensions of comparison} (e.g., methodology, application, dataset). The target paper is then compared to related publications along these \emph{same} extracted dimensions. This comparative analysis, inspired by tournament selection, yields quantitative novelty scores reflecting the relative improvement, equivalence, or inferiority of the target paper in specific aspects. By aggregating these scores and visualizing their progression, for instance, through dynamic evolution graphs and comparative radar charts, KnoVo facilitates researchers not only to assess originality and identify similar work, but also to track knowledge evolution along specific research dimensions, uncover research gaps, and explore cross-disciplinary connections. We demonstrate these capabilities through a detailed analysis of 20 diverse papers from multiple scientific fields and report on the performance of various open-source LLMs within the KnoVo framework.
\end{abstract}

\begin{keywords}
Research Novelty \sep Knowledge Evolution \sep Knowledge Representation \sep Large Language Models (LLMs) \sep Citation Network Analysis \sep Knowledge Graphs \sep Scientometrics
\end{keywords}

\maketitle

\section{Introduction}
\label{sec:introduction}

The accelerating pace of scientific publication constitutes a significant challenge to the entire research community: how to efficiently situate new ideas within the vast landscape of existing work to both understand their evolution and assess their novelty. Researchers aiming to build upon prior art, as well as reviewers and funding agencies evaluating new advancements, must determine whether an introduced work genuinely advances the state-of-the-art or simply re-treads familiar ground \cite{zhao2025review, amplayo2019evaluating, YanTZ20, foster2021surprise}. Traditional manual literature reviews, although beneficial, are intrinsically time-intensive, subjective, and progressively inadequate in addressing the rapid expansion of published literature. Existing quantitative metrics, such as citation counts and h-index, primarily measure historical impact, a lagging indicator that does not directly address the core question of an idea's novelty at its inception or its subsequent progression \cite{hou2022new, cohen2017should, Ivancovsky_Baror_Bar_2024}. Furthermore, while common similarity search methods can identify related publications, they lack the granularity to pinpoint \textit{how}, and on \textit{what specific dimensions}, a new contribution differs from previous work. This leaves researchers struggling to construct a clear, quantitative, dimension-specific, and temporally aware understanding of an idea's trajectory and true novelty, forcing them to manually sift through a large number of papers \cite{thelwall2022scopus}. These limitations highlight fundamental gaps in the tools available for knowledge discovery and evaluation, and motivate basic questions about how research progress is assessed. Specifically:

\begin{itemize}
    \item \emph{Can novelty be quantified?} Is it possible to create a quantitative, dimension-specific indicator of a research contribution's originality in comparison to the body of existing work that transcends subjective assessments and coarse-grained metrics? This calls for a methodology that can distinguish and contrast particular features of novelty rather than merely general similarity.
    
    \item \emph{Can the evolution of research ideas be tracked?} Knowing when a certain concept or method was first proposed and how it has been developed, enhanced, or replaced over time is essential to understanding the trajectory of scientific advancement. Is it possible to create tools that offer this temporal context?
    
    \item \emph{Can relevant prior work be efficiently identified?} In order to evaluate the novelty of a proposed work, researchers need a way to rapidly identify the papers that are most directly relevant, eliminate noise, and concentrate on important comparison dimensions.
\end{itemize}

\begin{figure*}[htbp]
    \centering
    \begin{subfigure}[b]{0.8\textwidth}
        \centering
        \includegraphics[width=\textwidth]{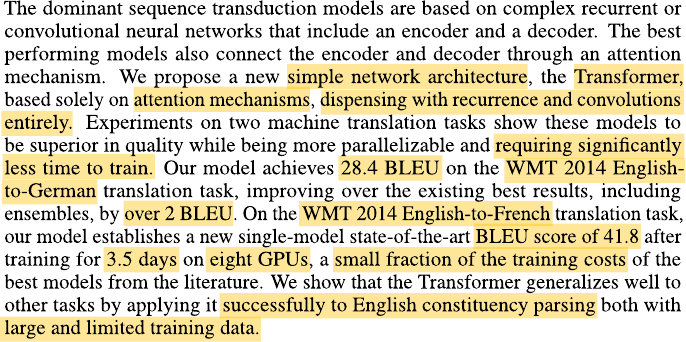}
        \caption{Target paper abstract ("Attention is All You Need" \cite{vaswani2017attention})}
    \end{subfigure}
    \par\medskip

    \begin{subfigure}[b]{0.8\textwidth}
        \centering
        \includegraphics[width=\textwidth]{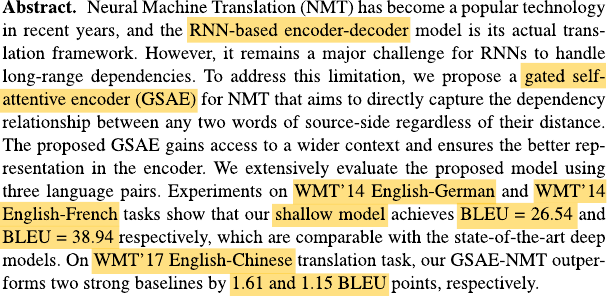}
        \caption{Related paper abstract ("Gated Self-attentive Encoder for Neural Machine Translation" \cite{10.1007/978-3-030-29551-6_58})}
        \label{fig:sub_related_abstract}
    \end{subfigure}
    \caption{Dimension and Value Extraction from Abstracts. This figure presents excerpts with highlighted phrases from (a) the target paper "Attention is All You Need" \cite{vaswani2017attention} and (b) a related paper \cite{10.1007/978-3-030-29551-6_58}, illustrating KnoVo's automated dimension extraction. KnoVo first dynamically identifies dimensions and their values from the target paper (a), exemplified by dimensions such as \texttt{Architecture Type} (value: \texttt{"Transformer"}), \texttt{Technique Used} (value: \texttt{"Attention Mechanism"}), and performance metrics like \texttt{English to German BLEU Score} (value: \texttt{"28.4 BLEU on WMT 2014 English-to-German"}). Subsequently, as shown in the related paper (b), KnoVo seeks out values for these \emph{same} target-derived dimensions to ensure a consistent feature space for comparison. This process enables direct quantitative comparison; for instance, on the \texttt{English-to-German} and \texttt{English-to-French} BLEU score dimensions, the target paper's scores (28.4 and 41.8, respectively) are superior to those of the related paper (26.54 and 38.94). Consequently, KnoVo's analysis would award the target paper positive scores for these dimensions, quantifying its advancement.}


    \label{fig:dimension_extraction_example}
\end{figure*}

In order to directly address these issues, this paper presents KnoVo (Knowledge Evolution), an intelligent system that automates crucial aspects of novelty assessment in a process inspired by human expert analysis. The fundamental idea behind KnoVo is that novelty is multifaceted and relative by nature; a paper is novel in relation to certain aspects of prior research rather than just being \emph{novel} or \emph{not novel} \cite{funk2017dynamic, bu2021multidimensional}. KnoVo's approach first identifies the key contributions of a target paper and then compares these same contributions against related work. For the crucial first step, KnoVo leverages the intrinsic capabilities of Large Language Models (LLMs). As modern LLMs are built upon \emph{attention} mechanisms, they are adept at identifying semantically important sections of text based on their vast training. KnoVo harnesses this power to have LLMs dynamically extract fine-grained \emph{dimensions of contribution} directly from a target paper's abstract, moving beyond methods that rely on pre-defined categories \cite{hofstra2020diversity}. This extraction process is illustrated in Figure~\ref{fig:dimension_extraction_example} with the seminal "Attention is All You Need" paper \cite{vaswani2017attention}, where KnoVo formalizes the authors' specific claims of novelty. Second, along these same dynamically extracted dimensions, KnoVo uses LLMs to compare the target paper to publications in its multi-layered citation network. This comparative analysis, inspired by tournament selection, yields a score (+1, 0, or -1) for each dimension, indicating whether the target paper improves upon, is equivalent to, or is superseded by related work.

The KnoVo system generates quantitative novelty scores for the target paper by aggregating these dimension-specific scores. Furthermore, as demonstrated in Figure~\ref{fig:time_series_individual}, KnoVo provides a dynamic view of knowledge progression by analyzing these scores across multiple papers and publication dates, enabling the visualization of the historical emergence and evolution of specific research dimensions.  Crucially, KnoVo is designed for accessibility and practical use. It avoids the expenses and dependencies of proprietary APIs by utilizing local, open-source LLMs and operating on readily available abstracts. This paper includes a contrastive analysis of several of these LLMs, such as Deepseek-r1 \cite{guo2025deepseek}, Gemma3 \cite{gemmateam2025gemma3technicalreport}, within the KnoVo framework. We demonstrate KnoVo's capabilities through a detailed analysis of \textbf{twenty diverse papers} from multiple scientific fields, assessing varying types and levels of contribution.


The main contributions of this paper are:

\begin{enumerate}

    \item \textbf{Conceptual Framework:} We introduce KnoVo, a novel framework for quantitative, automated evaluation of research novelty. Designed to be suitable even with little or no forward citation data, KnoVo sets itself apart by conducting a multi-dimensional comparison within a paper's citation network.
    
    \item \textbf{System Implementation:} We create a functional prototype of KnoVo, showcasing its primary capabilities. The method promotes accessibility and cost-efficiency by employing locally hosted, open-source LLMs for dynamic dimension extraction and comparison analysis.
    
    \item \textbf{Empirical Demonstration:} We provide a detailed analysis of twenty diverse papers, showcasing KnoVo's ability to quantify novelty, visualize score evolution across dimensions and time, and identify papers with similar novelty profiles. This analysis demonstrates KnoVo's comprehension of knowledge progression.

    \item \textbf{Methodological Advancements:} we describe new methods for dynamic dimension extraction from abstracts, LLM-driven comparative scoring, and a time-series approach to novelty tracking, providing a methodological foundation for future research in automated novelty assessment.
    
    \item \textbf{LLM Evaluation:} We present a comparative analysis of several open-source LLMs (such as Deepseek-r1 \cite{guo2025deepseek}, Gemma3 \cite{gemmateam2025gemma3technicalreport}, Llama3.2 \cite{dubey2024llama}, and Mistral Small \cite{mistral2025small31}) within the context of the KnoVo framework, evaluating their performance on the important tasks of dimension extraction and comparison and making recommendations for the model selection for this use case.
    
\end{enumerate}


\section{Related Work}

The assessment of research novelty has been addressed from various standpoints, spanning bibliometrics, information retrieval, and natural language processing \cite{foster2015tradition, WAGNER20191260}.  We organize related work into four main approaches: LLM-based bibliometric methods, knowledge graph-based evolution tracking, citation network analysis techniques, and semantic content analysis approaches. We discuss each category, emphasizing the strengths and weaknesses of standard methods and contrasting them with KnoVo.

\textbf{LLM-Based Bibliometric Approaches:} Recent work has explored the use of Large Language Models (LLMs) for novelty assessment \cite{shibayama2021measuring}. \cite{lin2024evaluatingenhancinglargelanguage} propose RAG-Novelty, a retrieval-augmented generation technique that simulates a peer review process using an LLM (e.g., GPT-4 \cite{openai2024report}) to judge a manuscript's novelty in the context of retrieved, topically similar papers. This approach leverages the reasoning capabilities of LLMs, and they show it outperforms other LLM prompting strategies.  While RAG-Novelty shares KnoVo's use of LLMs, it varies considerably in its approach. RAG-Novelty focuses on overall novelty judgment within a simulated peer review context, while KnoVo extracts specific dimensions of novelty and performs fine-grained comparisons. In contrast to RAG-Novelty's dependence on proprietary models like GPT-4 \cite{openai2024report}, KnoVo also stresses the use of local, open-source LLMs for accessibility and cost-effectiveness.  Furthermore, KnoVo incorporates a temporal dimension by analyzing both references and citations, allowing for the tracking of novelty evolution, a feature not present in RAG-Novelty.


\textbf{Knowledge Graph-Based Evolution Tracking:} These approaches use graph evolution to detect novelty and represent scientific knowledge as networks of entities and connections. \cite{AMPLAYO2018542} build multi-level knowledge graphs and quantify the structural disruption brought about by a new paper's release. A high reconstruction error in an autoencoder trained on the graph indicates higher novelty. \cite{hofstra2020diversity} count the number of previously unconnected concepts that are linked for the first time by a new publication. \cite{MARTINDEDIEGO2021101188} quantify novelty in medical research by measuring the divergence of a paper's concept graph from a pre-existing knowledge base. Though these methods effectively capture the structural novelty of linking concepts, they vary significantly from KnoVo. KnoVo does not rely on a pre-existing knowledge graph or ontology.  Instead, it dynamically extracts dimensions of novelty from the text itself, making it applicable to any field, regardless of the availability of structured knowledge representations. Furthermore, KnoVo provides a quantitative multi-dimensional novelty score, whereas knowledge graph approaches usually focus on discovering new connections or structural changes without necessarily quantifying the degree of novelty along specific dimensions. KnoVo's use of LLMs allows for a more nuanced understanding of the contribution's content in addition to its structural impact on a knowledge graph.

\textbf{Techniques for Citation Network Analysis:} Conventional bibliometric techniques use citation network structural patterns to assess novelty. For example, \cite{uzzi2013atypical} offered a metric based on the atypicality of reference combinations, demonstrating that high-impact papers typically use both traditional and unconventional sources. By measuring novelty by identifying first-time combinations of cited journals, \cite{WANG20171416} brought attention to a potential bias against novelty in science. Based on the structure of the citation network, these methods quantify novelty using metrics such as journal combinations and co-citation frequency. KnoVo differs greatly from these citation-based approaches. They are not content-agnostic; they only employ citation styles. KnoVo, on the other hand, analyzes the content of papers (via their abstracts) to extract and compare aspects of novelty. Citation-based methods are capable of identifying unusual pairings, but they cannot identify which specific study components are unique or why a combination is unique. KnoVo's multifaceted approach offers this more thorough analysis.  Furthermore, KnoVo considers both references and citations, providing a temporal perspective often absent from traditional citation analysis.

\textbf{Semantic Content Analysis Methods:} Using Natural Language Processing (NLP), semantic approaches examine paper text to identify novel concepts based on word usage, topic similarity, and contribution statements. \cite{chen2019automatic} extracts unique n-grams from paper abstracts to find potential novel ideas. \cite{JEON2023101450} uses word embeddings and outlier detection to identify papers with novel topics. \cite{Wang2023Novelty} analyzes contribution sentences using deep neural models and topic modeling to assess novelty in relation to earlier research. Both semantic approaches share KnoVo's emphasis on textual content, despite differences in comparison strategy and level of granularity.  While methods based on n-grams or topic modeling often provide a general evaluation of topical novelty, they may miss subtle variations in methodology or application. KnoVo, on the other hand, extracts specific contribution dimensions, allowing for a more sophisticated and accurate comparison. Many semantic approaches focus on finding new terms or topics, whereas KnoVo assesses improvement and relative contribution within a given dimension.


\section{Methodology}
\label{sec:methodology}

\begin{figure*}[htb]
  \centering
  \includegraphics[width=0.8\textwidth]{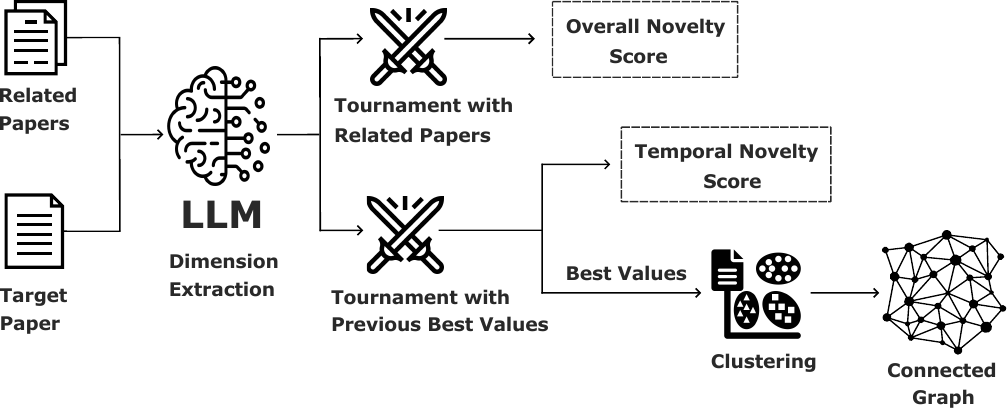}
    \caption{KnoVo system workflow diagram, illustrating key stages from dimension extraction to novelty analysis.}
  \label{fig:knovo_workflow}
\end{figure*}

The KnoVo system evaluates the novelty of a target paper \(T\) relative to a chronologically ordered set of related papers \(P = \{P_0, P_1, \dots, P_N\}\). The overall workflow of the KnoVo system, illustrating the key stages from data input to the final novelty analysis, is depicted in Figure~\ref{fig:knovo_workflow}. The following subsections break down each component of this process to detail its core operational mechanics.

\subsection{Dimension Extraction}
\label{subsec:dimension_extraction}

The first stage identifies key dimensions of contribution relevant to the target paper \(T\). We utilize a LLM (\(\Lambda_{\text{extract}}\)) guided by prompt \(P_{\text{extract}}\) applied to the target abstract \(A_T\). This yields a set of \(m\) dimensions \(D = \{d_1, \dots, d_m\}\) and their corresponding values \(V_T = \{v_{T,1}, \dots, v_{T,m} \}\) specific to \(T\). The prompt focuses on extracting specific, comparable features (e.g., architecture, technique, benchmark result).

\[
(D, V_T) = \Lambda_{\text{extract}}(A_T, P_{\text{extract}})
\]

Subsequently, using \(\Lambda_{\text{extract}}\) guided by the dimensions \(D\) derived from the target paper, the corresponding values \(A_{R_i}\) are extracted for each related paper \(R_i \in R\) to ensure a consistent feature space for comparison.


\subsection{Comparative Analysis}
\label{sec:comparative_analysis}

This stage defines the framework for comparing the target paper \(T\) against related papers \(R_i \in R\) along the extracted dimensions \(D\). Inspired by tournament selection, an LLM-based function, \(\Lambda_{\text{compare}}\), determines \(T\) relative to each \(R_i\) on every shared dimension \(d \in D\). This is achieved by comparing the target paper's value \(v_{T,d}\) with the related paper's value \(v_{R_i,d}\) for that dimension. The comparison yields a score \(S(d, R_i)\) representing relative improvement (\(+1\)), equivalence (\(0\)), or inferiority (\(-1\)), with the \(-1\) outcome being most directly applicable to numerical or quantitatively comparable dimensions (as justified in Sec.~\ref{sec:rationale_ternary_score}). Conceptually, this score is:

\[
S(d, R_i) =
\begin{cases}
  \text{null}, & \text{if dimension } d \text{ is not applicable/addressed in } R_i, \\
  0, & \text{if } T \text{ and } R_i \text{ are equivalent on } d, \\
  1, & \text{if } T \text{ is superior to } R_i \text{ on } d, \\
  -1, & \text{if } R_i \text{ is superior to } T \text{ on } d.
\end{cases}
\label{eq:score_conceptual}
\]

These dimension-specific scores across all related papers form the basis for the subsequent overall and temporal novelty calculations. Furthermore, for each comparison, \(\Lambda_{\text{compare}}\) also generates a textual justification, \(J(d, R_i)\), detailing the reasoning for its output, which is retained for logging and potential human oversight.

\subsection{Novelty Score Calculation}

KnoVo calculates two types of novelty scores: an overall novelty score and temporal novelty scores.

\subsubsection{Overall Novelty Score (\(\Omega\))}
\label{subsec:overall_novelty_knovo}

The overall novelty score (\(\Omega\)) aggregates the dimension-specific comparison scores \(S(d, R_i) \in \{1, 0, -1, \text{null}\}\), generated as described in Sec.~\ref{sec:comparative_analysis}, across all dimensions \(d \in D\) and all related papers \(R_i \in R\). These scores populate a \emph{Score Matrix}, denoted \(S_{\text{tournament}}\), which serves as the input for calculating \(\Omega\) as follows:

\textbf{Dimension Importance Weighting:} The relative importance (weight \(w_d\)) of each dimension \(d \in D\) is determined based on the comparison outcomes in \(S_{\text{tournament}}\), guided by the principle from information theory that higher uncertainty or variability within a system often correlates with higher information content (entropy). We adapt this principle to measure dimension importance by defining a dimension's \emph{Information Density} concerning the target paper \(T\)'s novelty. In our context, this density reflects the frequency with which \(T\) introduces advancements (\(S(d, R_i)=1\)) relative to the compared papers \(R_i\). The rationale is that dimensions exhibiting a higher proportion of +1 scores (\(P_d(1)\)) signify greater informative variance regarding \(T\)'s novel standing compared to related work; this higher frequency of observed advancement is interpreted as higher Information Density, indicating greater significance for that dimension in the overall novelty assessment. The raw weight \(w'_d\) is therefore set directly proportional to this measure, using the proportion \(P_d(1)\):


\[
w'_d = P_d(1) 
\label{eq:raw_weight_prop1} 
\]

These raw weights are then normalized to yield the final dimension weights \(w_d\) such that \(\sum_{d \in D} w_d = 1\):

\begin{gather*} 
w_d = \frac{w'_d}{\sum_{j=1}^{m} w'_j} \label{eq:normalized_weight_prop1} \\ 
\quad (\text{if } \sum_{j=1}^{m} w'_j > 0 \text{ else } w_d = 1/m), \quad \text{where } m=|D| 
\end{gather*}

This weighting calculation process is encapsulated within the function \(\Lambda_{\text{weight}}\).

\textbf{Dimension Score Calculation:} Next, for each dimension \(d\), we analyze the set \(S_{d}\) containing the numeric scores (\(1, 0, -1\)) from the pairwise comparisons \(\{ S(d, R_i) \mid R_i \in R \}\). Let \(N_d\) be the total non null count of these scores. We compute the proportion \(P_d(s)\) for each score value \(s \in \{1, 0, -1\}\) as follows (assuming \(N_d > 0\)):

\[
P_d(s) = \frac{\text{Count of score } s \text{ in } S_{d}}{N_d}, \quad \text{for } s \in \{1, 0, -1\}
\label{eq:proportion_calc}
\]

The score for dimension \(d\), denoted \(Score_d\), incorporates an equivalence weight \(\alpha\) (default 0.5) that controls the influence of contributions found to be equivalent (s=0) relative to target work: 

\[
Score_d = P_d(1) + (\alpha \cdot P_d(0)) - P_d(-1)
\]

If no valid scores exist for the dimension (\(N_d = 0\)), then \(Score_d = 0\).

\textbf{Final Weighted Aggregation:} The overall novelty score \(\Omega\) is the weighted sum of the individual dimension scores using the \(P_d(1)\)-derived weights \(w_d\):

\[
\Omega = \sum_{d \in D} w_d \cdot Score_d
\label{eq:overall_score_final} 
\]

This calculation provides a nuanced, aggregated measure of the target paper's novelty, weighted by the Information Density (frequency of improvements) observed across dimensions. Algorithm~\ref{alg:overall_score} provides the complete pseudocode for calculating \(\Omega\).

\begin{algorithm}[htb!]
\caption{KnoVo Overall Novelty Score (\(\Omega\))}
\label{alg:overall_score}
\begin{algorithmic}[1]
\Require Score Matrix \(S_{\text{tournament}}\); Dimension set \(D\); LLM \(\Lambda_{\text{weight}}\); Equivalence weight \(\alpha\).
\Ensure Overall Novelty Score \(\Omega\).

\State \(W \gets \Lambda_{\text{weight}}(D)\) 
\State \(\Omega \gets 0.0\)
\ForAll{\(d \in D\)}
    \State \(N_d \gets \text{count}(\{ S(d, R_i) \in S_{\text{tournament}} \mid S(d, R_i) \neq \text{null} \})\)
    \If{\(N_d > 0\)}
        \State \(P_d(1) \gets \text{count}(\{ S(d, R_i) = 1 \}) / N_d\)
        \State \(P_d(0) \gets \text{count}(\{ S(d, R_i) = 0 \}) / N_d\)
        \State \(P_d(-1) \gets \text{count}(\{ S(d, R_i) = -1 \}) / N_d\)
        \State \(Score_d \gets P_d(1) + (\alpha \cdot P_d(0)) - P_d(-1)\)
        \State \(\Omega \mathrel{+}= W.get(d, 0) \cdot Score_d\)
    \EndIf
\EndFor
\State \Return \(\Omega\)
\end{algorithmic}
\end{algorithm}

\subsubsection{Temporal Novelty Scores}
\label{subsec:temporal_novelty}

The Temporal novelty scores trace the evolution of novelty across dimensions over time. The set of related papers \(R\), included with the target paper T placed in its correct chronological position, is ordered by publication date. For each dimension \(d \in D\), we define a cumulative novelty score \(\nu(d,i)\) at the \(i\)-th paper in this ordered sequence to track the number of advancements relative to the best-so-far state observed up to that point for dimension \(d\).

\textbf{Initialization:} For every \(d \in D\),
\[
\nu(d,0) = 0.
\]

\textbf{Iteration:} For \(i = 1,2,\dots,|R|\), let \(R_i\) be the \(i\)-th paper in the chronologically ordered sequence (which includes \(T\)) with its extracted dimension values \(A_{R_i}\). The comparison function \(\Lambda_{\text{compare}}\) (as defined in Sec.~\ref{sec:comparative_analysis}) then evaluates \(R_i\) against the best value state observed up to the previous step for dimension \(d\), denoted \(\beta(d, i-1)\). For dimensions identified as categorical, \(\beta(d, i-1)\) retains the history of distinct best values encountered. Using a comparison prompt \(P_{\text{compare}}\), \(\Lambda_{\text{compare}}\) yields a score \(S(d, R_i) \in \{1, 0, -1, \text{null}\}\) for each dimension \(d\) and paper \(R_i\). These scores populate a Temporal Score Matrix, denoted \(S_{\text{temporal}}\), capturing the outcome of comparing each paper against the evolving best-so-far state for each dimension:

\[
S(d, R_i) = \Lambda_{\text{compare}}(A_{R_i}, \beta(d, i-1), P_{\text{compare}}, d)
\label{eq:lambda_compare_temporal}
\]

Then, for each \(d \in D\), the cumulative score \(\nu(d,i)\) is updated based on this comparison outcome:

\[
\nu(d,i) =
\begin{cases}
\nu(d,i-1) + 1, & \text{if } S(d, R_i) = 1,\\[1mm]
\nu(d,i-1), & \text{otherwise.}
\end{cases}
\]


An overall measure of cumulative novelty for each paper \(R_i\) is obtained by calculating the average cumulative novelty score, \(\overline{\nu}(i)\). This score averages the individual dimension scores \(\nu(d, i)\) across all dimensions \(d \in D\):
\[
\overline{\nu}(i) = \frac{1}{|D|} \sum_{d \in D} \nu(d, i)
\label{eq:average_temporal_score}
\]
where \(|D|\) is the number of dimensions. The score \(\overline{\nu}(i)\) represents the total combined advancement reflected in the research trajectory up to paper \(R_i\). To isolate the specific contribution introduced by paper \(R_i\), we define the Marginal Average Advancement, \(\Delta \overline{\nu}(i)\), as the change in the average cumulative score:
\[ 
\Delta \overline{\nu}(i) = \overline{\nu}(i) - \overline{\nu}(i-1) \quad (\text{for } i \ge 1) 
\label{eq:marginal_average_score} 
\]
This score \(\Delta \overline{\nu}(i)\) represents the average advancement across dimensions attributable specifically to paper \(R_i\). Algorithm~\ref{alg:timeseries} details the computation of the temporal novelty scores.

\begin{algorithm}[htb!]
\caption{Temporal Novelty Score Calculation}
\label{alg:timeseries}
\begin{algorithmic}[1]
\Require Ordered \(R\) with values \(A_{R_i}\), Dimensions \(D\), Comparison Logic \(P_{\text{compare}}\), Dimension Types Map.
\Ensure Time-series cumulative scores \(\{\nu(d,i)\}\), Average scores \(\{\overline{\nu}(i)\}\), Temporal Score Matrix \(S_{\text{temporal}}\).

\State Initialize \(\{\beta(d, 0) \mid d \in D\}\) 
\State Initialize \(S_{\text{temporal}}\) 
\State Initialize \(\nu(d,0) \gets 0, \forall d \in D\) 
\State \(\overline{\nu}(0) \gets 0\) 

\For{\(i=1\) to \(|R|\)}
    \State \(S_i \gets 0\) 
    \ForAll{\(d\in D\)}
        \State \(S(d, R_i) \gets \Lambda_{\text{compare}}(A_{R_i}, \beta(d, i-1), P_{\text{compare}}, d)\)
        \State Store \(S(d, R_i)\) in \(S_{\text{temporal}}[i, d]\) 
        \If{\(S(d,R_i)=1\)}
            \State \(\nu(d,i)\gets\nu(d,i-1)+1\)
            \State Let \(v_{i,d}\) be the value for dimension \(d\) from \(A_{R_i}\)
            \If{\(\text{IsCategorical}(d)\)}
                \State \(\beta(d, i) \gets \beta(d, i-1) \mathbin{\|} v_{i,d}\) 
            \Else 
                \State \(\beta(d, i) \gets v_{i,d}\)
            \EndIf
        \Else
            \State \(\nu(d,i)\gets\nu(d,i-1)\)
            \State \(\beta(d, i) \gets \beta(d, i-1)\) 
        \EndIf
        \State \(S_i \mathrel{+}= \nu(d,i)\) 
    \EndFor
    \State \(\overline{\nu}(i) \gets S_i / |D|\) 
\EndFor
\State \Return \(\{\nu(d,i)\}\), \(\{\overline{\nu}(i)\}\), \(S_{\text{temporal}}\)
\end{algorithmic}
\end{algorithm}

\subsection{Dimension-Specific Evolution Analysis}
\label{sec:dimension_evolution}

To provide a granular view of progress within a research area, KnoVo analyzes the evolution trajectory for individual dimensions \(d\). This involves tracking the cumulative number of advancements \(\nu(d,i)\) over the sequence of papers (derived from Sec.~\ref{subsec:temporal_novelty}) and modeling the interconnections between the specific papers \(R_i\) that introduced these advancements. An advancement by paper \(R_i\) in dimension \(d\) corresponds to \(\Delta \nu(d,i) = \nu(d,i) - \nu(d,i-1) = 1\) (or equivalently, \(S(d, R_i)=1\)). This subsequent analysis focuses on the subset of advancing papers \(P_d^+ = \{ R_i \in R \mid \Delta \nu(d,i) = 1 \}\), identified from the Temporal Score Matrix (\(S_{\text{temporal}}\)), to characterize these improvements and reveal the flow of ideas through clustering and relationship graph construction.

\subsubsection{Clustering Contributions}
\label{sec:value_clustering}

Advancements within the dimension $d$ are grouped based on semantic similarity to identify related concepts. Building on foundational ideas of learning distributed representations of concepts \cite{hinton1986learning}, the textual contribution values $\{v_{k,d}\}$ from the improving papers $P_k \in P_d^+$ are transformed into semantic vector embeddings $e_k$ using a pre-trained language model, $M_{embed}$:

\[
e_k = M_{embed}(v_{k,d})
\label{eq:embedding}
\]

These embeddings \(E = \{e_k\}\) are then processed using a density-based clustering algorithm \(A_{cluster}\) (e.g., DBSCAN) to partition the papers \(P_d^+\) into groups based on embedding similarity. Each paper \(P_k\) is assigned a cluster label \(c_k\) (including a potential ``Noise'' label):

\[
\{c_k\} = A_{cluster}(E)
\label{eq:clustering}
\]

This clustering helps identify distinct sub-themes or parallel approaches within the dimension's evolution.

\subsubsection{Constructing the Relationship Graph}
\label{sec:relationship_graph}

To understand the evolutionary pathways, a directed weighted graph \(G_d = (V_d, E_d)\) is constructed for the dimension. The nodes \(V_d\) correspond to the papers \(P_k \in P_d^+\). Edges \(E_d\) signify inferred conceptual or methodological links between pairs of papers \((P_i, P_j)\), directed from the earlier published paper to the later published paper. The weight \(w_{ij}\) of an edge reflects the \emph{confidence} in the relationship, determined by applying prioritized heuristics based on cluster membership, lexical similarity, and LLM assessment. Let \(c_k\) be the cluster label for paper \(P_k\) (where \(c_k=N\) signifies noise), and let \(\Lambda_{R}(i,j)\) be true if the LLM (\(\Lambda_{\text{relate}}\)) confirms relatedness between \(P_i\) and \(P_j\). The confidence \(w_{ij}\) is assigned based on the first matching condition in the following prioritized order:

\[
w_{ij} =
\begin{cases}
5, & \text{if } c_i=c_j \neq N \text{ and high lexical overlap of } v_{i,d}, v_{j,d} \\
4, & \text{if } c_i=c_j \neq N \text{ and low lexical overlap of } v_{i,d}, v_{j,d} \\
3, & \text{if } \Lambda_{R}(i,j) \land c_i \neq N \land c_j \neq N \\
2, & \text{if } \Lambda_{R}(i,j) \land (c_i=N \oplus c_j=N) \\
1, & \text{if } \Lambda_{R}(i,j) \land c_i=N \land c_j=N \\
0, & \text{otherwise}
\end{cases}
\label{eq:confidence_heuristic}
\]

where \(\oplus\) denotes the exclusive OR operation. Edges \(E_d\) are added to the graph connecting pairs of papers where \(w_{ij} > 0\), with \(w_{ij}\) representing the assigned confidence score. This graph models the potential interconnections and development sequence of ideas within the dimension. Algorithm~\ref{alg:dim_analysis} outlines the overall process.

\begin{algorithm}[htb!]
\caption{Dimension-Specific Analysis}
\label{alg:dim_analysis}
\begin{algorithmic}[1]
\Require Advancing papers set \(P_d^+\) (with values, dates); Models \(M_{embed}, \Lambda_{\text{relate}}\); Algorithm \(A_{cluster}\).
\Ensure Clustered papers \(\{P_k \text{ with } c_k\}\); Relationship graph \(G_d = (V_d, E_d)\).

\State \(E \gets \emptyset\) 
\ForAll{\(P_k \in P_d^+\)}
    \State \(e_k \gets M_{embed}(v_{k,d})\) 
    \State Add \(e_k\) to \(E\)
\EndFor
\State \(\{c_k\} \gets A_{cluster}(E)\) 

\State \(V_d \gets P_d^+\) 
\State \(E_d \gets \emptyset\) 
\ForAll{pair \((P_i, P_j) \in P_d^+ \times P_d^+, i \neq j\)}
    \State \(conf \gets w_{ij}\) 
    \If{\(conf > 0\)}
        \State Add edge \((P_i, P_j)\)\textsubscript{directed by date} weight \(conf\) to \(E_d\)
    \EndIf
\EndFor
\State \Return \(\{P_k \text{ with } c_k\}\), \(G_d = (V_d, E_d)\)
\end{algorithmic}
\end{algorithm}


\subsubsection{Temporal Forest Construction}
\label{sec:temporal_forest}

While the heuristic graph \(G_d\) (Sec.~\ref{sec:relationship_graph}) models potential connections, identifying the primary evolutionary pathways requires extracting the most significant structural links. To achieve this, we construct a Temporal Evolution Forest, inspired by progressive alignment methods and Maximum Spanning Forest algorithms (e.g., Kruskal's).

This method selects the most significant, chronologically valid edges between the advancing papers \(P_d^+\). Potential directed edges \((P_i, P_j)\) are considered only if \(Year(P_i) \le Year(P_j)\). Let \(\Delta t_{ij} = Year(P_j) - Year(P_i)\) represent the non-negative year gap between paper \(P_j\) and paper \(P_i\). Edges are scored using a function \(\sigma_{\text{edge}}(P_i, P_j)\) based on both their connection confidence and this temporal proximity \(\Delta t_{ij}\). The connection confidence is given by the heuristic weight \(w_{ij}\) derived using the process detailed in Section~\ref{sec:relationship_graph} (Eq.~\ref{eq:confidence_heuristic}). The edge score, \(\sigma_{\text{edge}}(P_i, P_j)\), combines confidence \(w_{ij}\) with the year gap \(\Delta t_{ij}\), prioritizing high confidence and small gaps (low \(\Delta t_{ij}\)):

\[
\sigma_{\text{edge}}(P_i, P_j) = \frac{w_{ij}^{\gamma}}{(\Delta t_{ij} + 1)^{\delta}}
\label{eq:edge_score}
\]

where \(\gamma \ge 0\) and \(\delta \ge 0\) control the emphasis on confidence versus proximity (e.g., using \(\gamma=1, \delta=1\) balances both). Edges are only considered if their confidence score \(w_{ij} > 0\).

The forest \(G'_d = (V'_d, E'_d)\), where \(V'_d = P_d^+\), is then built using a greedy approach detailed in Algorithm~\ref{alg:temporal_forest}. This algorithm sorts potential edges by \(\sigma_{\text{edge}}\) and iteratively adds the highest-scoring edge that connects two previously disconnected components (checked via Union-Find), ensuring no cycles are formed. The resulting directed forest \(G'_d\) represents the inferred primary pathways of idea evolution within dimension \(d\).

\begin{algorithm}[htb!]
\caption{Temporal Evolution Forest Construction}
\label{alg:temporal_forest}
\begin{algorithmic}[1]
\Require Advancing papers set \(P_d^+\);  Scoring parameters \(\gamma, \delta\).
\Ensure Temporal Evolution Forest edge set \(E'_d\).

\State Initialize list \(PotentialEdges\)
\ForAll{pair \((P_i, P_j) \in P_d^+ \times P_d^+, i \neq j\)}
    \If{\(Year(P_i) \le Year(P_j)\)}
        \State \(conf \gets w_{ij}\)
        \If{\(conf > 0\)}
            \State \(score \gets \sigma_{\text{edge}}(P_i, P_j)\) 
            \State Add edge \((P_i, P_j)\) with \(score\) to \(PotentialEdges\)
        \EndIf
    \EndIf
\EndFor

\State Sort \(PotentialEdges\) by \(score\) descending
\State Initialize Union-Find structure \(UF\) with nodes from \(P_d^+\)
\State \(E'_d \gets \emptyset\)

\ForAll{edge \((P_i, P_j)\) with \(score\) in sorted \(PotentialEdges\)}
    \If{\(UF.union(P_i, P_j)\)}
        \State Add directed edge \((P_i, P_j)\) to \(E'_d\)
    \EndIf
\EndFor
\State \Return \(E'_d\)
\end{algorithmic}
\end{algorithm}

\begin{figure*}[htb]
  \centering
  \includegraphics[width=1\textwidth]{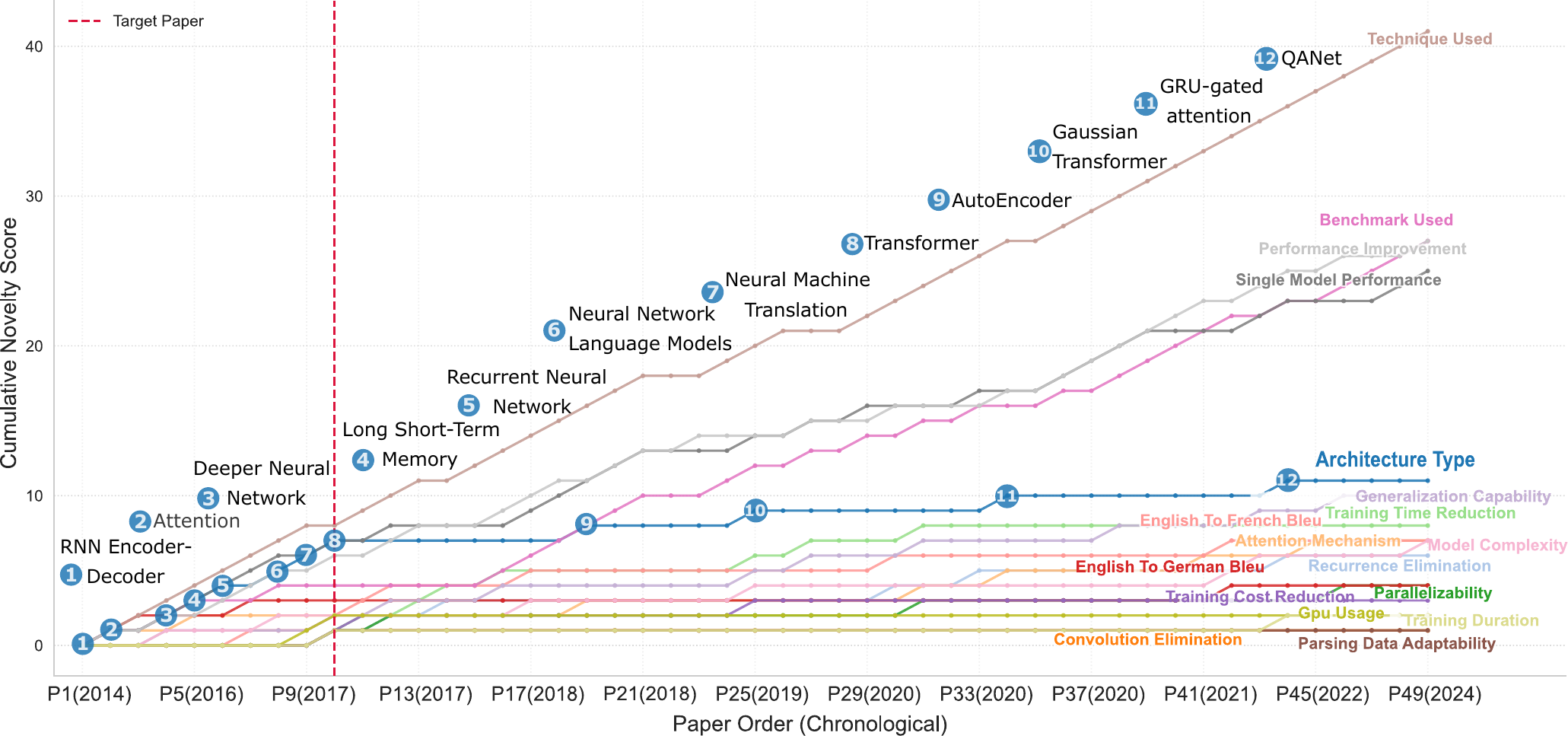}
    \caption{Cumulative Novelty Score Evolution by Dimension. Time-series analysis of cumulative novelty scores for key research dimensions, based on methods from \cite{vaswani2017attention}. (a) Each line tracks cumulative novelty score (y-axis) per dimension (legend) over time (x-axis, chronological paper order). Points mark the dimension's state-of-the-art score upon publication. (b) The dashed red line indicates the target paper's publication time and scores. Papers to the left are prior work/references; those to the right are subsequent publications. (c) Upward jumps indicate novel contributions to a dimension by a paper; flat segments represent periods with no recorded advancements relative to the prior state-of-the-art.}
  \label{fig:time_series_individual}
\end{figure*}

\subsection{Design Rationale}
\label{sec:design_rationale}

The design of KnoVo resulted from an iterative process, balancing hypotheses about effective novelty evaluation with experimental findings and practical implementation considerations. This section elaborates on the rationale behind key methodological decisions (e.g., the use of a ternary scoring system \({+1}, 0, -1\)).

\subsubsection{Target-Centric Fixed Dimensions}
\label{sec:rationale_fixed_dimensions}

A major design choice in KnoVo is the extraction of comparison dimensions \(D\) only from the target paper \(T\) (Sec.~\ref{subsec:dimension_extraction}), ensuring these dimensions offer a solid basis, or fixed axiom, for all comparisons involving \(T\). This uniformity is necessary for effective evaluation. Without such a defined dimension set, comparing relative superiority becomes unstable and potentially intransitive. For instance, if we find that paper A improves on B relative to dimensions \(D_{AB}\) (denoted \(A >_{D_{AB}} B\)), and subsequently find \(B\) improves on C relative to a different set of dimensions \(D_{BC}\) (\(B >_{D_{BC}} C\) where \(D_{AB} \neq D_{BC}\)), we lack the common frame of reference required to infer any relationship between A and C. The concept of superiority is particular to the dimensions investigated (\(D_{AB}\) vs. \(D_{BC}\)), rendering transitive reasoning invalid. Such instability hinders accurate aggregation of pairwise comparisons into an overall score (like \(\Omega\), Sec.~\ref{subsec:overall_novelty_knovo}) and invalidates consistent temporal monitoring of advancements within a dimension (like \(\nu(d,i)\), Sec.~\ref{subsec:temporal_novelty}). While user-specified dimensions might be provided, depending only on them would divorce the assessment from the paper's fundamental claims; This technique assures the analysis is founded in the target paper's key contributions.


\subsubsection{Ternary Comparison Score}
\label{sec:rationale_ternary_score}

KnoVo's pairwise comparison score \(S(d, R_i)\)  (Sec.~\ref{sec:comparative_analysis}) draws inspiration from tournament selection paradigms, where relative results are typically more relevant than absolute fitness values.  When developing the output codomain for the comparison function \(\Lambda_{\text{compare}}\), we investigated possibilities such as a in-depth 5-point scale (e.g., \(\{+1, +0.5, 0, -0.5, -1\}\)).  However, we ultimately selected the simpler ternary codomain \(\{+1, 0, -1\}\) conceptually signifying improvement, equivalence, or inferiority.

 This decision prioritizes assessment robustness and reproducibility, particularly given the nature of LLM-based evaluation.  Eliciting consistent, objective magnitude scores (such +0.5 vs +1) from LLMs is tough due to task complexity and potential subjectivity.  Furthermore, while judging superiority (+1) or equivalence (0) is generally feasible, reliably determining strict inferiority (-1) for categorical dimensions (e.g., deeming one conceptual approach definitively worse than another without specific quantitative backing) proved particularly prone to subjective LLM interpretation and inconsistency during development.  Therefore, while the scoring framework conceptually incorporates \(-1\) (as defined in Sec.~\ref{sec:comparative_analysis}), its practical applicability and dependability are primarily centered on numerical or other quantitatively comparable aspects.  Constraining the primary judgment, especially for categorical types, mostly to separating novelty/improvement (+1) from equivalence (0) gives a more steady and interpretable signal.  The overall significance of developments is then addressed through dimension weighting (\(w_d\)).

\subsubsection{Best-So-Far Comparison}
\label{sec:rationale_best_so_far}

A primary goal of KnoVo's temporal analysis is to identify when novel approaches or significant advancements emerge within a dimension \(d\) over time, allowing these key contributions to be highlighted and analyzed (Sec.~\ref{sec:dimension_evolution}). To achieve this efficiently, we employ a "best-so-far" comparison strategy. This approach compares the contribution value \(v_{i,d}\) of each incoming paper \(R_i\) against the optimal state achieved by all prior papers in the sequence, denoted \(\beta(d, i-1)\). Conceptually, this is analogous to tracking the maximum value encountered so far in a numerical sequence or maintaining the set of unique best concepts seen for categorical dimensions. An advancement is formally registered (\(S(d, R_i)=1\)), thus incrementing the cumulative score \(\nu(d,i)\), only when the current paper's value \(v_{i,d}\) is measured by \(\Lambda_{\text{compare}}\) as superior to the prior best state represented by \(\beta(d, i-1)\). By focusing solely on contributions that surpass the previous state-of-the-art, this strategy effectively isolates the specific papers \(P_d^+\) responsible for pushing the dimension frontier forward within the dimension.

\subsubsection{Marginal Advancement Score}
\label{sec:marginal_score}

To determine the specific impact of individual papers within the temporal sequence, we draw inspiration from methods measuring rates of change, analogous to using gradients or derivatives to understand instantaneous impact in optimization processes. While the average cumulative score \(\overline{\nu}(i)\) effectively tracks the total accumulated advancement along the research trajectory, it doesn't isolate the contribution attributable solely to paper \(R_i\). For instance, consider two papers \(R_j\) and \(R_i\) in the sequence where \(j < i\). It is possible for \(\overline{\nu}(i) > \overline{\nu}(j)\) even if paper \(R_i\) introduced less novelty (\(\Delta \overline{\nu}(i) \approx 0\)) than the earlier paper \(R_j\) (\(\Delta \overline{\nu}(j) > 0\)), simply because \(\overline{\nu}(i)\) inherently incorporates advancements from papers published between steps \(j\) and \(i\). Therefore, to quantify the advancement attributable specifically to paper \(R_i\), we defined the Marginal Average Advancement \(\Delta \overline{\nu}(i)\) (Sec.~\ref{subsec:temporal_novelty}). This metric, calculated as the difference \(\overline{\nu}(i) - \overline{\nu}(i-1)\), provides a score suitable for comparing the distinct contributions of individual papers across the chronological sequence.

\begin{figure}[htbp]
  \centering
  \includegraphics[width=1\textwidth]{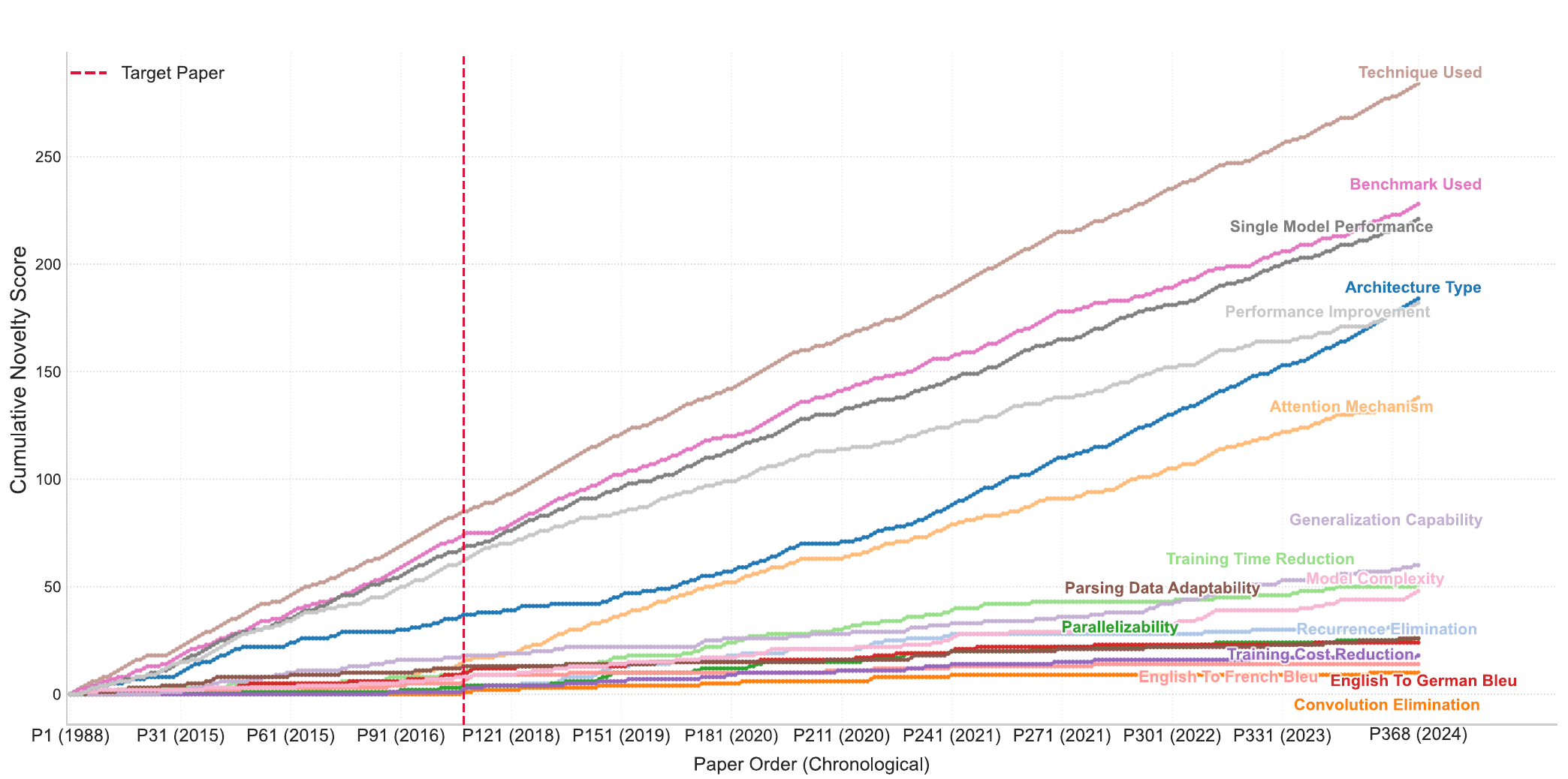}
 \caption{Cumulative Novelty Score Evolution by Dimension using the 2-layer citation network data constructed for the target paper \cite{vaswani2017attention} (approx. 368 papers). This plot illustrates the denser score trajectories compared to Figure~\ref{fig:time_series_individual} (which uses the 1-layer citation network data for the same target paper). Each line tracks the cumulative novelty score \(\nu(d,i)\) (y-axis) per dimension over the chronological paper order (x-axis). Upward jumps indicate novel contributions (\(S(d, R_i)=1\)) relative to the prior state-of-the-art for that dimension.}
  \label{fig:app_cumulative_dense}
\end{figure}

\subsubsection{Clustering and Relationship Heuristics}
\label{sec:rationale_clustering_heuristics}

While tracking cumulative novelty scores (as in Figure~\ref{fig:time_series_individual}) is effective for sparse networks, these visualizations become dense and difficult to interpret as the citation network deepens. As illustrated in Figure~\ref{fig:app_cumulative_dense}, analyzing a 2-layer network with hundreds of papers makes it intractable to visually discern distinct evolutionary pathways or trace the origins of key ideas from the plot alone. To deconstruct this complexity and model the flow of ideas, KnoVo adopts a structured analysis inspired by methodologies in biological sequence analysis \cite{Durbin_Eddy_Krogh_Mitchison_1998}. The process first uses semantic clustering (Sec.~\ref{sec:value_clustering}) to group papers that introduced similar advancements ($P_d^+$). Subsequently, it constructs a relationship graph ($G_d$) using defined heuristics (Sec.~\ref{sec:relationship_graph}) to model the progression between these conceptual clusters, drawing inspiration from how probabilistic models are used to understand state transitions in gene classification. This defined, two-stage process was deliberately chosen over using an LLM to directly classify papers into ambiguous categories like "seminal" or "incremental". Such a direct approach is problematic as it relies on opaque LLM knowledge, lacks consistent rules for relative ranking (e.g., ensuring "supportive" Paper B is indeed between "seminal" Paper A and "incremental" Paper C), and suffers from poor reproducibility. In contrast, KnoVo's heuristic approach provides a transparent, multi-faceted, and reproducible method for inferring the connections that form the evolutionary pathways of scientific ideas.

\subsection{Prompt Engineering}
\label{subsec:prompt_engineering}

The reliability and precision of KnoVo's analysis depend on sophisticated prompt engineering to guide LLMs through multi-step reasoning tasks. Our approach involves crafting distinct, structured prompts for each core function: dynamic dimension extraction, fixed-dimension value extraction, and comparative analysis. This strategy, inspired by how a human expert first identifies key aspects of a target paper and then seeks out those same aspects in related work for comparison, is operationalized through carefully designed LLM interactions.

For the initial dynamic dimension extraction ($\Lambda_{\text{extract}}$), the LLM is prompted to act as a specialized research assistant. It follows detailed guidelines to identify specific contributions and formalize them into generalizable "dimension: value" pairs, avoiding overly broad terms while ensuring the resulting dimension keys are understandable and comparable across papers.

For subsequent steps, we shift to a more constrained approach using LLM function calling \cite{openai_function_calling} to ensure structured, reliable outputs. To extract values for the now-fixed dimensions from related papers, the LLM is instructed to call a function that populates a predefined JSON schema. This forces the model to return values only for the target-derived dimensions and to use an empty string for any dimension where a value is not found, ensuring a consistent feature space. Similarly, for the comparative analysis ($\Lambda_{\text{compare}}$), the LLM is provided with the dimension name, the target value, and the value to compare against, along with a detailed set of rules for assigning a score (+1, 0, or -1). It must return its result via a function call that yields both the score and a brief justification.

This methodology of combining expert personas, rule-based instructions, and a strong reliance on function calling for schema enforcement is central to KnoVo. It makes the LLM's analytical process more transparent, repeatable, and machine-readable, transforming it from a purely generative model into a predictable reasoning component within the KnoVo system.

\subsection{Implementation}
\label{sec:implementation}

The KnoVo system is implemented in Python, leveraging several key libraries for data manipulation, asynchronous processing, and interaction with Large Language Models (LLMs). We use \texttt{pandas} for data handling and structuring, enabling efficient processing of paper metadata and comparison results. \texttt{asyncio} and the \texttt{aiohttp} library (used internally by \texttt{ollama}) facilitate asynchronous communication with the LLMs, allowing for concurrent processing of multiple papers and dimensions, significantly improving performance. The \texttt{ollama} library provides a convenient interface for interacting with locally running, open-source LLMs. We also utilize \texttt{re} for regular expression matching, \texttt{requests} for basic HTTP requests, and \texttt{nest\_asyncio} to enable nested event loops in certain environments (e.g., Jupyter notebooks). Visualizations were generated using \texttt{matplotlib} and \texttt{plotly}.

The core evaluation presented in this paper were primarily performed using Gemma 3 large language models (e.g., \texttt{gemma3:12b}, \texttt{gemma3:27b}), leveraging their capabilities for structured data generation and comparative reasoning. Other open-source models, including Gemma 2, Deepseek-r1:7b, Llama3.2, and mistral-small, were also utilized during development or for specific auxiliary tasks like graph heuristic generation. Specific model versions, hyperparameters (typically temperature 0 for deterministic outputs). 

\section{Evaluation}

To determine the feasibility and potential of the KnoVo framework, we conducted experiments applying it to a curated dataset using selected open-source LLMs. This evaluation examines KnoVo's key analytical outputs—including novelty scores (\(\Omega\)), temporal evolution patterns (\(\nu(d,i)\), \(\overline{\nu}(i)\)), and dimension-specific idea evolution graphs—alongside its computational performance. The results demonstrate KnoVo's core functionality and potential for automated novelty assessment.

\subsection{Experimental Setup}
\label{sec:exp_setup}

This section details the dataset curated for our experiments and the selection process for the Large Language Models (LLMs) used to implement the KnoVo analysis pipeline.

\subsubsection{Dataset}
\label{sec:dataset}

To rigorously evaluate KnoVo, we curated a dataset comprising 20 target papers, selected according to a structured strategy emphasizing disciplinary diversity and varied contribution types. The selection spans multiple fields including Computer Science (with sub-fields such as Machine Learning/NLP and Databases/Systems), Biology/Medicine, Physics/Quantum Computing, Economics/Social Science, and Environmental Science, ensuring a broad testbed for KnoVo's generalizability. For each target paper, we constructed a 2-layer citation network. This involved collecting its immediate references and, recursively, the references of those references. In the forward direction, we gathered its immediate citations, with a maximum of 50 papers, and then subsequently collected up to 50 citations for each of these first-layer citing papers to form the second layer.

The data, including abstracts, metadata, and citation links, was primarily gathered using the Semantic Scholar API \cite{ammar-etal-2018-construction}. The choice of Semantic Scholar was deliberate, offering robust API access essential for large-scale data collection. While alternatives like DBLP \cite{DBLP:conf/www/Berners-Lee11} lack comprehensive metadata such as abstracts and publication dates, and often rely on Semantic Scholar for citation data, \cite{googlescholar} does not provide official API support, hindering systematic collection. A key advantage of the Semantic Scholar API is its \textit{relevance} sort feature for retrieving citations; although direct chronological sorting of citations can be limited, relevance-based sorting provided a temporally more balanced distribution of citing papers, crucial for analyzing evolutionary trends. The dataset includes a rich set of metadata fields for each paper, as detailed in Table~\ref{tab:dataset_columns}.

\begin{table}[htbp]
\centering
\caption{Key Metadata Fields in the Curated Dataset}
\label{tab:dataset_columns}
\begin{tabular}{l|c}
\toprule
\textbf{Column Name} & \textbf{Description} \\
\midrule
paperId & Unique Semantic Scholar identifier \\
title & Full title of the paper \\
abstract & Paper's abstract \\
authors & List of author names \\
publicationVenue & Name of the publication venue \\
year & Publication year \\
referenceCount & Number of references made \\
citationCount & Total number of citations received \\
influentialCitationCount & Count of influential citations \\
isOpenAccess & Boolean indicating open access status \\
openAccessPdf & URL to open access PDF, if available \\
fieldsOfStudy & List of identified fields of study \\
publicationDate & Full publication date, if available \\
journal & Journal details (name, volume, pages) \\
type & Relationship (citation, reference) \\
layer & Network layer in our collection \\
\bottomrule
\end{tabular}
\end{table}

The availability of structured metadata, such as "authors", "publicationVenue", and "fieldsOfStudy", facilitates a more in-depth analysis of individual papers and their interconnected networks, offering deeper insights into the scholarly landscape. Key characteristics of the overall curated dataset of 20 target papers and their networks are summarized in Table~\ref{tab:dataset_with_scores}.

\subsubsection{Model Selection}
\label{sec:model_selection}

In line with KnoVo's design principles of accessibility and cost-effectiveness, we focused on open-source, locally runnable Large Language Models (LLMs) for the core components of our system. This approach avoids the substantial expense and potential rate limits associated with proprietary API-based models like GPT-4 \cite{openai2024gpt4technicalreport} or Claude \cite{anthropic2024claude3}, making KnoVo more readily deployable and scalable, particularly given the numerous LLM calls required for dimension extraction and comparison across the citation network.

We explored a range of locally runnable open-source models, including variants of Llama3.2 \cite{dubey2024llama}, Gemma2 \cite{team2024gemma}, Deepseek-r1 \cite{guo2025deepseek}, Gemma 3 \cite{gemmateam2025gemma3technicalreport}, and Mistral Small \cite{mistral2025small31} (see Table~\ref{tab:model_specs}). While initial tests showed Gemma 2 offered good stability, Gemma 3 models were ultimately selected as the primary engine for KnoVo due to providing the most consistent and reproducible structured outputs across multiple runs, a critical factor for our pipeline, compared to other models that exhibited more variability. As a newer generation model, Gemma 3 also offered advanced capabilities. We adopted a strategy of assigning models based on task complexity: the larger \texttt{gemma3:27b} variant was utilized for the most demanding reasoning tasks—initial dynamic dimension extraction ($\Lambda_{\text{extract}}$, Sec.~\ref{subsec:dimension_extraction}) and comparative analysis ($\Lambda_{\text{compare}}$, Sec.~\ref{sec:comparative_analysis})—owing to its strong logical capabilities. However, given memory constraints observed with the base 27B model, quantized versions (e.g., \texttt{gemma3:27b-it-qat}) were sometimes necessary. Less demanding auxiliary tasks, like fixed value extraction or type determination, were often handled by the smaller \texttt{gemma3:12b} variant. For certain, high-volume procedures, the \texttt{mistral-small} model (24B parameters) was used to balance computational cost and performance.  For instance, \texttt{mistral-small} was substantially faster than Gemma 3 while evaluating graph connection heuristics (Sec.~\ref{sec:relationship_graph}), a procedure that evaluates several possible connections between clusters. This makes it the more sensible option for this task.

\begin{table}[htbp]
\centering
\caption{LLM Model Specifications}
\label{tab:model_specs}
\begin{tabular}{l|c|c|c}
\toprule
Model  & Parameters & Context Length & Quantization \\
\midrule
Llama3.2 \cite{dubey2024llama} & 3B & 131072 & Q4\_K\_M \\
Gemma2 \cite{team2024gemma} & 9B & 8192 & Q4\_0 \\
Deepseek-r1 \cite{guo2025deepseek} & 7B & 131072 & Q4\_K\_M \\
Gemma3  \cite{gemmateam2025gemma3technicalreport} & 27B,12B & 131072 & Q4\_K\_M \\
Mistral Small \cite{mistral2025small31} & 24B & 32768 & Q4\_K\_M \\
\bottomrule
\end{tabular}
\end{table}

\begin{figure}[htbp]
\centering
\includegraphics[width=0.8\linewidth]{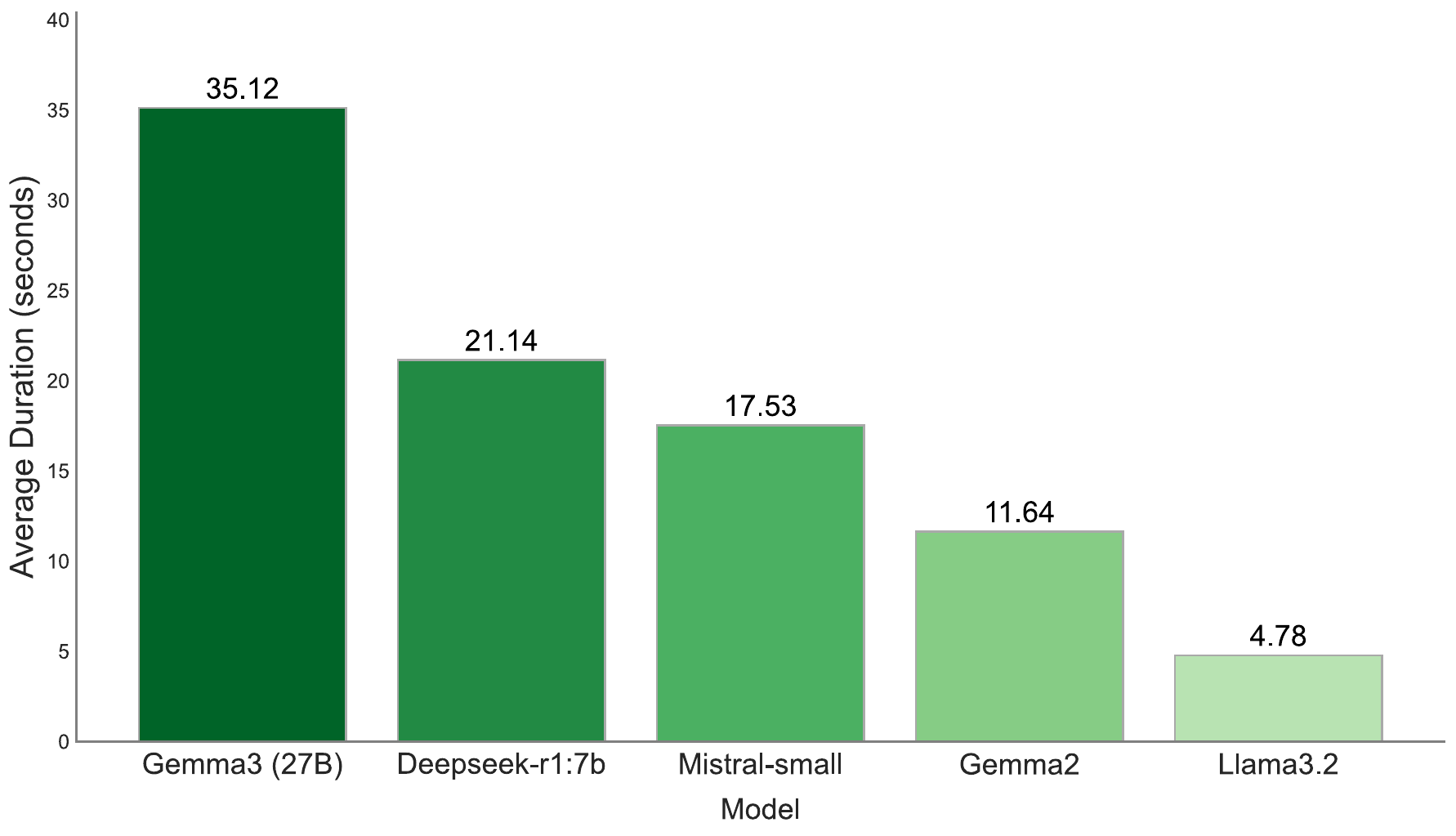} 
\caption{Average time (seconds) for the initial dynamic dimension extraction ($\Lambda_{\text{extract}}$) per abstract using various LLMs. Measurements were performed on a Windows machine equipped with 1 NVIDIA A6000 GPU.}
\label{fig:extraction_time}
\end{figure}

\subsection{KnoVo Analysis Outputs}
\label{sec:knovo_outputs}

Building upon the experimental setup, this subsection details the primary analytical outputs generated by KnoVo for the evaluated target papers. These include the dynamically extracted dimensions, the calculated overall (\(\Omega\)) and temporal (\(\nu(d,i)\), \(\overline{\nu}(i)\)) novelty scores, and the resulting dimension-specific idea evolution graphs derived from clustering and MST analysis. To maintain a clear narrative flow and provide a detailed demonstration across these different outputs, we will primarily use the target paper "Attention is All You Need" \cite{vaswani2017attention} as a running example throughout the following subsubsections.

\subsubsection{Dynamic Dimension Extraction Results}
\label{sec:eval_dim_extraction}

A core component of KnoVo is the dynamic extraction of key dimensions of contribution from a target paper's. These dimensions, representing specific aspects of novelty claimed by the authors, form the basis for subsequent comparative analysis. Rather than relying on predefined categories, KnoVo uses an LLM ($\Lambda_{\text{extract}}$) – primarily \texttt{gemma3:27b} for this task – guided by prompt \(P_{\text{extract}}\) to identify and extract these dimensions directly from the text. This process yields a set of dimension-value pairs (\(D, V_T\)), where each dimension represents a distinct aspect of the paper's contribution, and the value provides a concise description. To illustrate the nature of these extracted dimensions, the following list shows a subset of the dimensions generated by \(\Lambda_{\text{extract}}\) for the target paper \cite{vaswani2017attention}:

\begin{itemize}
    \item \textbf{Architecture Type:} Transformer 
    \item \textbf{Technique Used:} Attention Mechanism 
    \item \textbf{Parallelizability:} Increased Parallelizability 
    \item \textbf{Training Time Reduction:} Significant Training Time Reduction 
    \item \textbf{English to German BLEU:} 28.4 BLEU on WMT 2014 English-to-German 
    \item \textbf{Model Complexity:} Simplified Network Architecture 
    \item \textbf{Performance Improvement:} Over 2 BLEU improvement over existing ensembles 
\end{itemize}

KnoVo's dynamic extraction process thus identifies a diverse range of author-specified contributions, capturing both qualitative advancements and quantitative metrics. Crucially, this extracted set of dimensions \(D\) serves as a fixed axiom (as discussed in Sec.~\ref{sec:rationale_fixed_dimensions}) for evaluating the target paper against related work. This focus on specific, comparable dimensions derived directly from the target paper—rather than relying on broad predefined categories—is fundamental to enabling the subsequent comparative analysis.

\subsubsection{Overall and Temporal Novelty Scores}
\label{sec:eval_scores}

KnoVo produces two key types of scores to assess novelty: the overall novelty score (\(\Omega\)) for the target paper relative to its network, and the average cumulative novelty score (\(\overline{\nu}(i)\)) tracking aggregate progress over the temporal sequence.

The overall novelty score \(\Omega\) (calculated as detailed in Sec.~\ref{subsec:overall_novelty_knovo}) provides a single value summarizing the target paper's novelty. Table~\ref{tab:dataset_with_scores} presents the calculated \(\Omega\) scores for the ten target papers in our dataset, alongside their basic characteristics. To illustrate the meaning of these scores, consider two examples from the table: the "Semantic program alignment..." paper \cite{10.1145/3314221.3314596} has \(\Omega = 0.33\), indicating a moderate level of advancement over its related work based on the extracted dimensions. In contrast, "Attention is Not All You Need..." \cite{vaswani2017attention} achieves \(\Omega = 0.97\), suggesting a significantly higher degree of novelty relative to its comparators across the dimensions KnoVo identified. These scores offer a quantitative basis for comparing paper originality, providing researchers with a rapid assessment of a work's standing relative to existing literature.

\begin{figure}[htbp]
\centering
\includegraphics[width=0.9\linewidth]{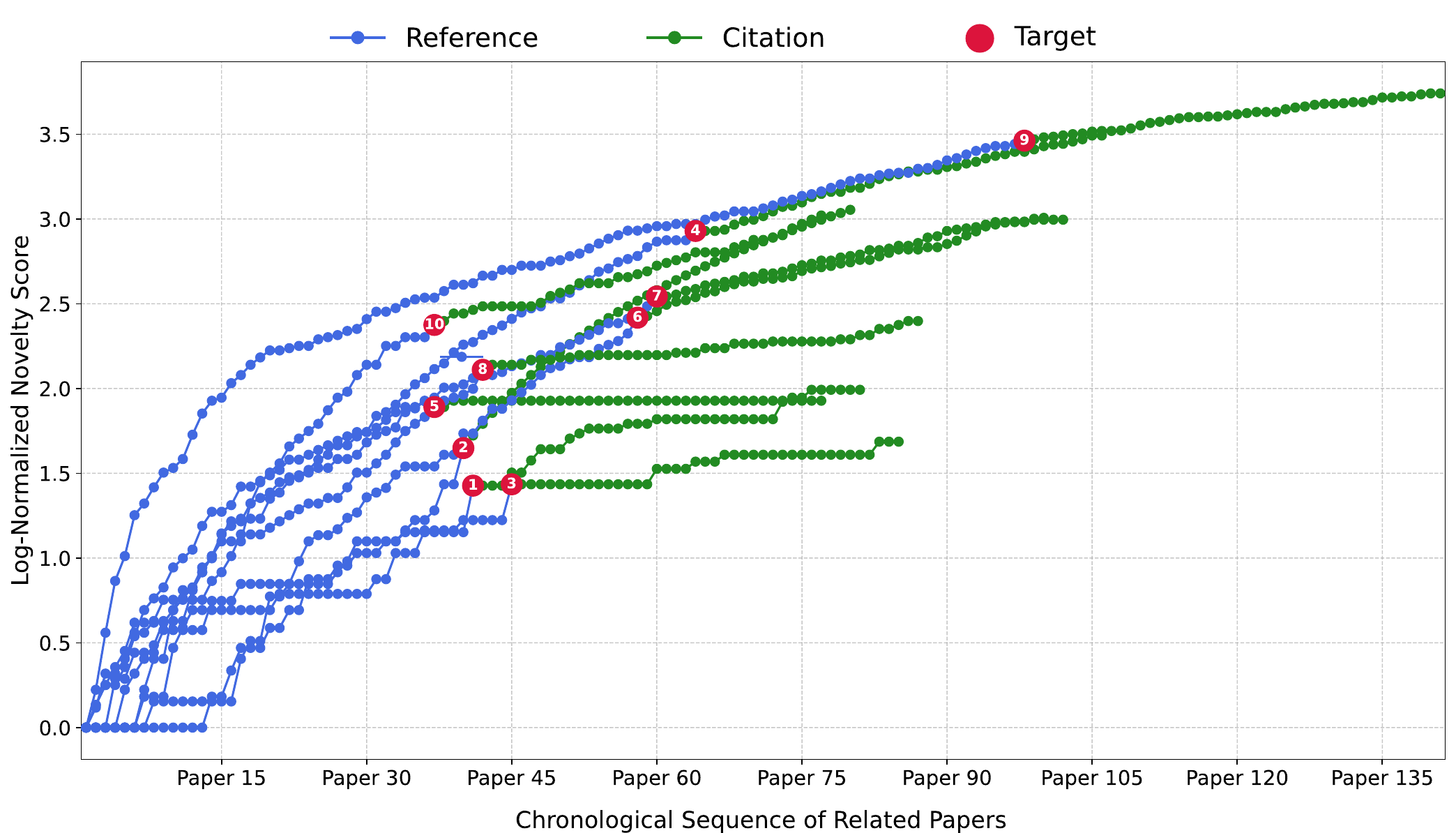}
\caption{This figure presents a time-series analysis of the combined, log-normalized novelty scores for ten target papers and their related literature (references in blue, citations in green). The large red dot on each paper's timeline indicates its overall novelty score and position. The x-axis represents the chronological sequence of related papers, and the y-axis shows the log-normalized novelty score (ln(1 + average dimension score)). Upward trends indicate the accumulation of novelty over time.}
\label{fig:time_series_overall}
\end{figure}

Complementing the static overall score, KnoVo also tracks the aggregate temporal evolution of novelty. Figure~\ref{fig:time_series_overall} visualizes this progression across all dimensions for each target paper's network. The y-axis plots the log-normalized average cumulative score, calculated as \(\ln(1 + \overline{\nu}(i))\), where \(\overline{\nu}(i)\) is the average score defined in Eq.~\ref{eq:average_temporal_score}. This logarithmic transformation is applied specifically for visualization to dampen the effect of large initial scores and provide a clearer view of incremental improvements over the entire sequence. The resulting plot allows for comparison of overall knowledge accumulation trends across different research trajectories, highlighting periods of rapid advancement or stagnation. The relative paper order on the x-axis, combined with year labels, aids in identifying key periods of innovation within each trajectory.

\subsubsection{Dimension-Specific Temporal Evolution}
\label{sec:eval_temporal_dim}

Beyond the aggregate trend KnoVo allows for examining the evolution trajectory within individual dimensions. Figure~\ref{fig:time_series_individual} illustrates this for the target paper \cite{vaswani2017attention}, plotting the cumulative novelty score \(\nu(d,i)\) for several key dimensions using the 1-layer citation network data. An upward step signifies that \(\Lambda_{\text{compare}}\) identified paper \(R_i\)'s contribution as novel (\(S(d, R_i)=1\)) for that dimension, meaning it either surpassed the numerical best-so-far value or introduced a categorical concept distinct from the history accumulated in \(\beta(d, i-1)\). 

This visualization reveals insights such as the rate of progress per dimension. Furthermore, Figure~\ref{fig:time_series_individual} annotates key advancements with the corresponding contribution values (\(v_{k,d}\)) identified by KnoVo. For instance, following the "Architecture Type" dimension trace, we see advancements labeled with concepts such as "RNN Encoder-Decoder" (Paper 1, 2014) and "Transformer" (Paper 25, 2017), highlighting the specific ideas recognized as novel at those points.

While this 1-layer visualization (Figure~\ref{fig:time_series_individual}) provides clarity for early progressions, analyzing the denser 2-layer citation network data (Figure~\ref{fig:app_cumulative_dense}) highlights a challenge. For dimensions where numerous related variations emerge over time (e.g., the many subtypes of GANs or Transformer-based models like BERT), the overlapping cumulative score lines make it difficult to distinguish specific evolutionary paths and conceptual relationships directly from the plot. These closely related but incrementally different ideas can become obscured in the density. This difficulty in interpreting dense trajectories motivates the subsequent analysis step. To better understand the flow of ideas and relationships between specific advancements (\(P_d^+\)), particularly in dimensions with rich evolutionary histories, we apply clustering and graph construction techniques, as detailed in the following subsection.

\begin{figure}[htb]
\centering
\includegraphics[width=0.7\linewidth]{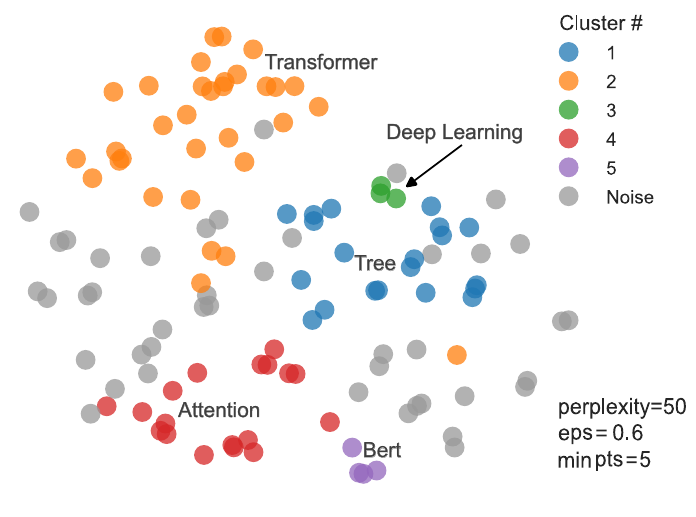}
\caption{t-SNE 2D projection visualizing semantic clusters of contribution values (\(v_{k,d}\)) from papers (\(P_d^+\)) that introduced advancements within an example dimension \(d\) (e.g., "architecture type"). Points are colored based on cluster assignments determined by DBSCAN applied to high-dimensional semantic embeddings (\(e_k\)) of the contribution values. t-SNE was used for dimensionality reduction. Manually added labels suggest interpretations for prominent clusters (e.g., "Transformer", "Attention"). Grey points indicate noise as identified by DBSCAN. This visualization aids in identifying distinct conceptual sub-themes among advancements.}
\label{fig:tsne_clusters} 
\end{figure}

\subsubsection{Case Study: Idea Evolution within a Dimension}
\label{sec:eval_idea_evolution}

To demonstrate KnoVo's capability for in-depth analysis of idea progression, we present a case study focusing on the "Architecture Type" dimension for the target paper \cite{vaswani2017attention}, utilizing the 2-layer citation network data. While temporal plots show advancements, interpreting the complex relationships within denser citation networks (e.g., involving hundreds of papers as shown in Figure~\ref{fig:app_cumulative_dense}) directly from these plots becomes difficult, thus necessitating further structural analysis.

First, applying the clustering technique described in Section~\ref{sec:value_clustering}, we group the papers \(P_d^+\) that introduced advancements in "Architecture Type" based on the semantic similarity of their contribution values (\(v_{k,d}\)). Figure~\ref{fig:tsne_clusters} visualizes these clusters via t-SNE projection, revealing distinct groupings corresponding to concepts like "Transformer", "Attention", "RNN", etc.

\begin{table}[htbp]
    \centering
    \caption{Evolution Forest Roots (Dimension: Architecture Types).}
    \label{tab:arch_roots}
    \setlength{\tabcolsep}{4pt}
    \begin{tabular}{r|c|l} 
        \toprule
        Root ID & Year & Value / Concept \\
        \midrule
        2    & 1993 & tree-structured \\
        10   & 2015 & convolutional neural network \\
        11   & 2015 & encoder-decoder \\
        13   & 2015 & deep neural networks (DNN) \\
        50   & 2020 & decoder-based \\
        61   & 2021 & BERT \\
        62   & 2021 & neural network \\
        63   & 2021 & gaussian embedder \\
        89   & 2023 & U-net \\
        \bottomrule
    \end{tabular}
\end{table}

\begin{figure}[htb]
\centering
\includegraphics[width=0.8\linewidth]{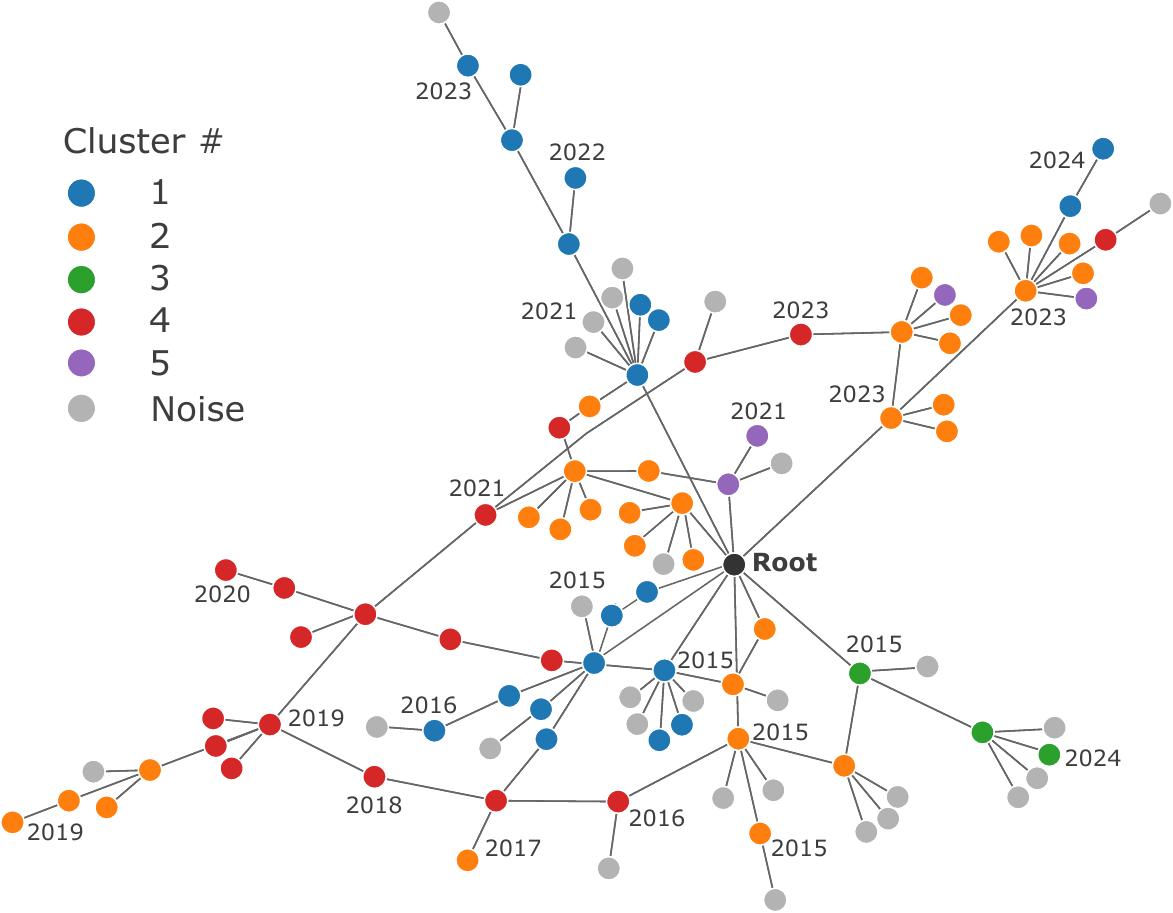}
\caption{Temporal evolution graph illustrating inferred developmental pathways for the "Architecture Types" dimension. Nodes represent papers identified as introducing architectural advancements (\(P_d^+\)), labeled by publication year and colored according to the semantic cluster of their contribution value (\(v_{k,d}\)), as determined by the analysis in Sec.~\ref{sec:value_clustering}. Directed edges depict the primary evolutionary links calculated by the temporal forest algorithm, representing high-confidence, temporally proximate connections. The layout, emanating from a central root, visually organizes these advancements to reveal branches and the flow of architectural ideas over time.}
\label{fig:arch_evolution_graph}
\end{figure}

Next, to extract the primary evolutionary pathways from the potential interconnections between these clustered advancements, we constructed the Temporal Evolution Forest using the method detailed in Section~\ref{sec:temporal_forest}. This algorithm selects the most significant chronological links based on connection confidence and temporal proximity. Figure~\ref{fig:arch_evolution_graph} displays the resulting directed forest for the "Architecture Types" dimension. Nodes are colored according to their semantic cluster (Figure~\ref{fig:tsne_clusters}), and the edges represent the inferred primary developmental timeline. This structure visually traces lineages, for example, showing paths originating from earlier concepts like "Encoder-Decoder" and leading towards "Attention" and "Transformer" based architectures. Table~\ref{tab:arch_roots} lists the root nodes identified within this Temporal Evolution Forest structure. These nodes, characterized by having no incoming forest edges, represent the inferred starting points of distinct evolutionary threads for "Architecture Type", often corresponding to the prominent semantic clusters seen in Figure~\ref{fig:tsne_clusters}.

\begin{figure}[htbp]
\centering
\includegraphics[width=0.9\linewidth]{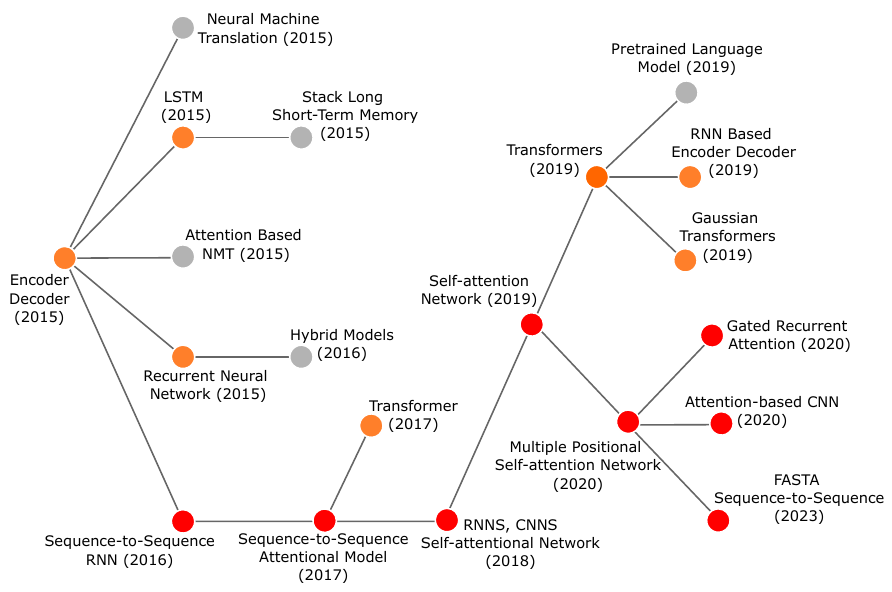}
\caption{Evolutionary pathway subgraph for the "Architecture Types" dimension, extracted from the full temporal graph (Figure~\ref{fig:arch_evolution_graph}). Nodes represent advancing papers/concepts (\(P_d^+\)) labeled by value (\(v_{k,d}\)) and year. Edges depict the primary developmental lineage determined by the temporal forest algorithm (Sec.~\ref{sec:relationship_graph}), illustrating the innovation trajectory (e.g., from Encoder-Decoder to Transformer variants).}
\label{fig:arch_evolution_subgraph}
\end{figure}

For a more detailed examination, Figure~\ref{fig:arch_evolution_subgraph} presents a subgraph extracted from the full temporal graph (Figure~\ref{fig:arch_evolution_graph}), focusing on the evolutionary lineage originating from the root concept "Encoder-Decoder" (node 11 in Table~\ref{tab:arch_roots}). This detailed view highlights the specific developmental pathway identified by KnoVo stemming from this earlier architectural approach, illustrating how subsequent related advancements connect and diverge over time. By combining clustering with the MST-based forest construction, KnoVo moves beyond simple timelines to model and visualize the structured evolution of specific research ideas.

\subsection{Validation}
\label{sec:validation}

To validate KnoVo's ability to quantify research novelty, we applied it to the curated dataset of 20 diverse target papers, analyzing their 2-layer citation networks as described in Section~\ref{sec:dataset}. Papers within these networks lacking abstracts were excluded from the comparative analysis. Key characteristics of the target papers, along with their processed citation network sizes, the number of dimensions KnoVo dynamically extracted, and their calculated KnoVo Overall Novelty Scores, are presented in Table~\ref{tab:dataset_with_scores}.

The scores in Table~\ref{tab:dataset_with_scores} demonstrate KnoVo's capacity to generate quantitative measures of novelty. The table includes the main Overall Novelty Score ($\Omega$), which assesses novelty within the broader analyzed citation network, and a "Novelty Score (Refs Only)." This latter score, calculated by comparing the target paper exclusively against its own references network, aims to capture its novelty relative to the existing literature at its inception; a fresh idea is expected to achieve a high score against its direct antecedents. Our qualitative review of KnoVo's outputs for these 20 papers—including the extracted dimensions, the pairwise comparisons, and the LLM-generated justifications—indicates that the system's assessments generally align with an expert understanding of these papers' contributions relative to their respective citation networks. For instance, for Paper 5 ("MapReduce: Simplified Data Processing on Large Clusters"), which received an $\Omega$ score of $0.6363$, KnoVo's lower score on a specific dimension like "fault tolerance" (value: "Automatic handling of machine failures") was justified by comparisons to cited works employing more robust mechanisms (e.g., "Byzantine Agreement"). Such justifications provide transparency and support the interpretation of the novelty scores. These findings suggest that KnoVo offers a meaningful, network-contextualized approach to novelty assessment, complemented by explainable reasoning. Further longitudinal and task-oriented validation is planned as outlined in our Future Work.

\begin{table*}[htbp]
\centering
\footnotesize
\setlength{\tabcolsep}{3pt}
\caption{Target Paper Characteristics and Calculated KnoVo Novelty Scores}
\label{tab:dataset_with_scores}
\begin{tabularx}{\textwidth}{@{} c >{\setlength{\leftskip}{5pt}\RaggedRight\arraybackslash}X >{\centering\arraybackslash}m{2.4cm} rrrr r @{}}
\toprule
\textbf{\#} & \textbf{Target Paper Title} & \textbf{Field} & \textbf{Orig. Cites} & \textbf{Network Size} & \textbf{KnoVo Dims} & \textbf{Novelty Score} & \textbf{Novelty Score (Refs)} \\
\midrule
1 & Attention is All you Need \cite{vaswani2023attentionneed} & \multirow{3}{*}{\parbox{2.2cm}{\centering Machine Learning / NLP (CS)}} & 121464 & 447 & 18 & 0.8936 & 0.9434 \\
2 & BERT: Pre-training of Deep Bidirectional Transformers... \cite{devlin2019bertpretrainingdeepbidirectional} &  & 89546 & 692 & 13 & 0.9646 & 0.9823 \\
3 & Zero-shot Generalizable Incremental Learning for Vision-Language... \cite{deng2024zeroshotgeneralizableincrementallearning} &  & 4 & 90 & 12 & 0.9069 & 0.8615 \\
\midrule
4 & Dynamo: Amazon's Highly Available Key-value Store \cite{10.1145/1294261.1294281} & \multirow{4}{*}{\parbox{2.2cm}{\centering Databases / Systems (CS)}} & 4522 & 209 & 10 & 0.9798 & 0.9803 \\

5 & MapReduce: Simplified Data Processing on Large Clusters \cite{10.1145/1327452.1327492} &  & 25772 & 148 & 11 & 0.6363 & 0.6000 \\
6 & DuckDB: An Embeddable Analytical Database \cite{10.1145/3299869.3320212} &  & 217 & 74 & 11 & 0.8733 & 0.7489 \\
7 & Apache Arrow DataFusion: A Fast, Embeddable, Modular... \cite{10.1145/3626246.3653368} &  & 5 & 20 & 14 & 0.9507 & 0.9713 \\
\midrule
8 & Initial Sequencing and Analysis of the Human Genome \cite{Lander2001InitialSA} & \multirow{4}{*}{\parbox{2.2cm}{\centering Biology / Medicine}} & 132447 & 916 & 9 & 0.8166 & 0.8647 \\
9 & A Programmable Dual-RNA–Guided DNA Endonu-clease... \cite{Jinek2012APD} &  & 13270 & 1403 & 10 & 0.9344 & 0.9668 \\
10 & Real-time Forecasts of the 2019-nCoV Epidemic in China... \cite{Roosa2020RealtimeFO} &  & 670 & 380 & 13 & 0.8081 & 0.7696 \\

11 & Global Burden of 369 Diseases and Injuries... (\cite{Christopher2020GlobalBO}) &  & 8978 & 1658 & 16 & 0.8578 & 0.9053 \\

\midrule
12 & Fault Tolerant Quantum Computation by Anyons \cite{Kitaev1997FaultTQ} & \multirow{4}{*}{\parbox{2.2cm}{\centering Physics / Quantum Computing}} & 5006 & 426 & 7 & 0.9703 & 1.0000 \\
13 & Quantum Supremacy using a Programmable Superconduct-ing... \cite{Arute2019QuantumSU} &  & 6291 & 653 & 11 & 0.9739 & 0.9753 \\
14 & Quantum Algorithms for Quan-tum Field Theories \cite{Jordan2011QuantumAF} &  & 473 & 499 & 10 & 0.9100 & 0.9225 \\

15 & Discretizing Quantum Field Theories for Quantum Simula-tion \cite{Farrelly2020DiscretizingQF} &  & 19 & 266 & 11 & 0.7700 & 0.7937 \\
\midrule
16 & The Central Role of the Propen-sity Score in Observational Studies... \cite{Rosenbaum1983TheCR} & \multirow{3}{*}{\parbox{2.2cm}{\centering Economics / Social Science}} & 29860 & 706 & 10 & 0.8260 & N/A \\
17 & Mostly Harmless Econometrics: An Empiricist's Companion \cite{Angrist2008MostlyHE} &  & 12684 & 414 & 10 & 0.9473 & 0.9727 \\
18 & Machine Learning: An Applied Econometric Approach \cite{Mullainathan2017MachineLA} &  & 1384 & 841 & 9 & 0.7206 & 0.7256 \\
\midrule
19 & Estimating Consumer Exposure to PFOS and PFOA \cite{Trudel2008EstimatingCE} & \multirow{2}{*}{\parbox{2.2cm}{\centering Chemical Science}} & 461 & 573 & 13 & 0.9069 & 0.8615 \\
20 & Thirty Years of Medical Surveil-lance in Perfluooctanoic Acid... \cite{Costa2009ThirtyYO} &  & 216 & 612 & 13 & 0.9090 & 0.9445 \\
\bottomrule
\end{tabularx}
\end{table*}

\subsection{Computational Performance}
\label{sec:computational_performance}

KnoVo's practical applicability depends on its computational performance, particularly when analyzing complex 2-layer citation networks. We measured execution times for key KnoVo operations on a \textit{Windows machine equipped with 1 NVIDIA A6000 GPU}, utilizing locally deployed open-source Large Language Models (LLMs). Table~\ref{tab:component_timings} details these performance metrics for a representative 2-layer network analysis, outlining the KnoVo functions and LLMs involved.

\begin{table}[htbp]
\centering
\small
\caption{Computational Time for Key KnoVo Operations on a Representative 2-Layer Citation Network}
\label{tab:component_timings}
\begin{tabular}{l|c|c|c}
\toprule
\textbf{KnoVo Operation} & \textbf{Core Function(s)} & \textbf{Avg. Time (s)} & \textbf{LLM Model Used} \\
\midrule
Initial Dimension Extraction & $\Lambda_{\text{extract}}$ & 37.42 & gemma3:27b \\
Related Paper Value Extraction & $\Lambda_{\text{extract}}$ & 2915.90 & gemma3:27b \\
Overall Novelty Comparisons & $\Lambda_{\text{compare}}$ & 9099.66 & gemma3:12b \\
Temporal Novelty Comparisons & $\Lambda_{\text{compare}}$ & 5420.79 & gemma3:12b \\
Evolution Graph Construction & $\Lambda_{\text{relate}}$ & 628.07 & mistral-small \\
\bottomrule
\end{tabular}
\end{table}

Table~\ref{tab:component_timings} reveals varied computational costs across KnoVo's pipeline. While initial dynamic dimension extraction from the target paper is swift (37.42s), operations involving extensive LLM interactions over a large 2-layer network are more demanding. For example, extracting values for fixed dimensions from approximately 450 related paper abstracts required 2915.90s (around 48 minutes), and the comprehensive pairwise comparisons for overall novelty took 9099.66s (over 2.5 hours). The construction of evolution graphs, leveraging $\Lambda_{\text{relate}}$ for heuristics alongside graph algorithms, also contributes significantly (628.07s).

Although a full KnoVo analysis on a deep network is computationally intensive, these timings confirm its feasibility on appropriately equipped local hardware. The results establish a performance baseline, and, as discussed in Section~\ref{sec:discussion}, future work targeting scalability through optimizations such as enhanced asynchronous processing, LLM call batching, caching, and a potential pre-computed knowledge graph backend will further improve KnoVo's efficiency.

\section{Applications}

\begin{figure}[ht!]
\centering
\includegraphics[width=0.9\linewidth]{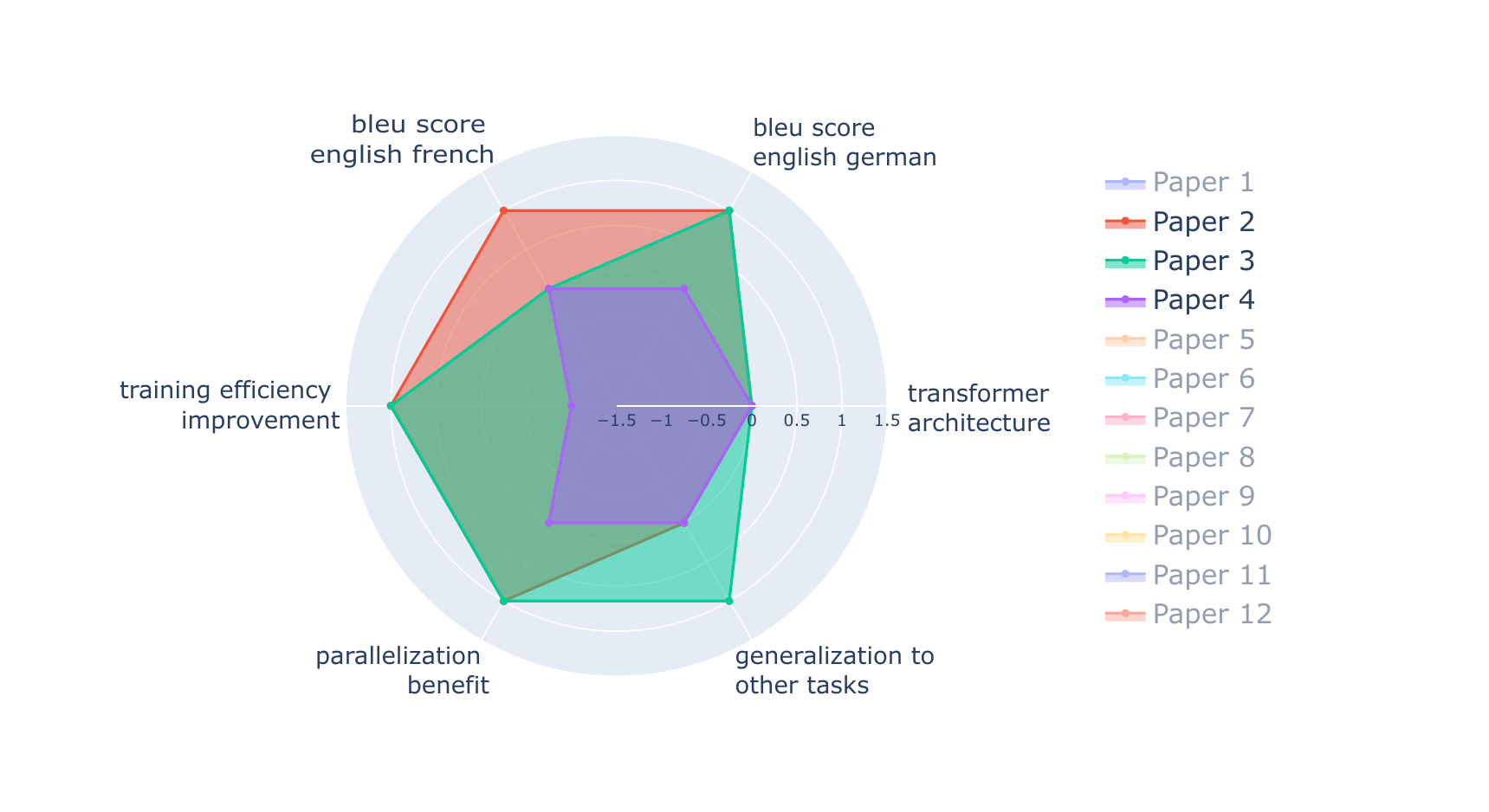} 
\caption{Radar Chart for Multi-Dimensional Novelty Comparison of papers related to \cite{vaswani2023attentionneed}. Vertices are extracted dimensions; polygons represent related papers, with radial distance indicating novelty scores ($S(d,R_i)$) on each dimension relative to the target.}
\label{fig:radar_chart}
\end{figure}

The KnoVo system, by quantifying novelty and visualizing research evolution across dynamically extracted dimensions, transcends traditional literature navigation, offering transformative applications for researchers, funding bodies, and the broader scientific endeavor. It moves beyond simple metrics by providing tools to understand the intricate fabric of scientific progress. A key visualization, the multi-dimensional radar chart (Figure~\ref{fig:radar_chart}), exemplifies this by offering an immediate comparative snapshot of how related papers score against a target paper (e.g., \cite{vaswani2023attentionneed}) along various dimensions of contribution. Each vertex represents a dimension, and a paper's novelty profile is captured by its polygon's shape and radial extent, allowing for rapid identification of its specific innovative strengths and areas of similarity or divergence from peers. For instance, as depicted, one paper might show broad novelty (larger polygon), while another excels along a specific dimensional axis. The versatility of KnoVo's framework enables a wide range of impactful use cases across different stakeholders (researchers, reviewers, bibliometric analysis, cross--disciplinary Investigation) in the research ecosystem, each benefiting from its ability to quantify, contextualize, and visualize novelty in scientific contributions.

\textbf{For Researchers.}
KnoVo serves as more than a search tool; it's an analytical environment akin to an enhanced, dynamic version of scholarly databases. It allows researchers to not only find papers but to understand their entire intellectual neighborhood, including an array of statistical insights and network connections. This aligns with the human research process: identifying a core idea and then exploring its context, antecedents, and impact. KnoVo significantly aids in the crucial but often arduous task of identifying research gaps. Traditional methods make it difficult to see what \textit{hasn't} been done or what was tried and abandoned prematurely. For example, a chemist modifying materials (e.g., X, Y) for a new drug can use KnoVo to analyze the evolution of research along material-specific dimensions, quickly identifying alternative materials that were explored but perhaps not pursued vigorously, or whose potential was overlooked due to the sheer volume of subsequent "noisy" publications. This capability can redirect research efforts towards more innovative paths, preventing duplication and illuminating neglected avenues—such as revisiting promising but "stalled" ideas from a previous research paradigm (e.g., late 2010s deep reinforcement learning concepts for current LLM development). KnoVo's temporal novelty scores ($v(d,i)$, $\overline{v}(i)$ from Sec 3.3.2) and dimension-specific evolution graphs ($G_d, G_d'$ constructed via Algorithms 3 and 4) are pivotal for this deep exploration.

\textbf{For Reviewers.}
Beyond individual research, KnoVo offers a more objective, quantitative foundation for evaluating the novelty of research proposals, augmenting expert judgment with data-driven insights derived from its comparative analysis ($\Lambda_{\text{compare}}$, Sec 3.2). This can foster more informed funding decisions and consistent evaluation standards. Agencies can utilize KnoVo for strategic portfolio analysis, pinpointing areas of high innovation, identifying emerging trends, and addressing potential gaps in their supported research, thereby also alleviating reviewer burden through partial automation of novelty assessment.

\textbf{For Bibliometric Analysis.}
KnoVo offers a more nuanced approach to analyzing scientific progress than traditional citation-based metrics. It goes beyond simple citation counts to consider the specific dimensions of novelty and how they evolve over time. This enables dynamic field mapping, revealing the emergence, convergence, and divergence of research areas. KnoVo can also help identify influential papers, not just based on their overall impact (citation volume), but also on their specific contributions to novelty within particular research dimensions, providing a richer understanding of scholarly influence.

\textbf{For Cross--Disciplinary Investigation.}
A significant hurdle in modern research is bridging deep expertise across disparate fields; for instance, an environmental scientist may not be an expert in chemical physics, making it challenging to uncover critical interdisciplinary connections. KnoVo's power is particularly pronounced here, excelling as an investigative tool for such complex, cross-disciplinary questions. Consider the challenge of understanding links between environmental factors and public health, such as the impact of chemicals like Perfluorooctanoic Acid (PFOA or C8). KnoVo can trace research histories, map studied effects, and identify commonalities or divergences across disparate studies (e.g., linking PFOA to thyroid, testicular, and liver cancers). By constructing dimension-specific evolution graphs ($G_d'$) for dimensions like "chemicals analyzed" or "observed health effects", and leveraging KnoVo’s clustering and LLM-informed relationship scoring, researchers can uncover subtle connections—such as common precursors or shared biological pathways across non-citing studies—that might take years to find manually. This can accelerate discoveries, for example, in identifying contamination sources (e.g., PFOA in Idaho drinking water, previously an unregulated and unmonitored area \cite{Costa2009ThirtyYO}) or assessing the environmental impact of industrial activities like deep-sea mineral extraction. KnoVo's ability to process diverse document types beyond academic papers, including regulatory reports or medical records, further amplifies this by enabling a more comprehensive understanding of complex issues.

To illustrate these investigative capabilities, a concrete application from our diverse dataset involved KnoVo analyzing "Thirty Years of Medical Surveillance in Perfluooctanoic Acid Production Workers"\cite{Costa2009ThirtyYO}. KnoVo extracted key dimensions and values, such as \texttt{study duration: 30 years}, \textit{population studied: PFOA production workers}, \textit{chemicals analyzed: Perfluooctanoic Acid (PFOA)}, and critical findings like \textit{cholesterol correlation: Significant association with PFOA levels} and \textit{metabolic interference: Probable interference with intermediate metabolism}. Analysis of this paper's citation network identified recurring themes, for instance, pinpointing multiple studies linking PFAS exposure to dyslipidemia, uric acid disruption, and specific cancer risks in occupational settings. The target PFOA paper itself achieved an Overall Novelty Score ($\Omega$) of $0.9090$ relative to its network, quantifying its significant contribution.

Moreover, KnoVo’s dimension-specific evolution analysis (Sec 3.4) facilitates the construction of dynamic knowledge graphs. For an important dimension like \textit{chemicals analyzed}, the system traces its evolution. These graphs are enriched by linking other dimensions from the same paper to relevant nodes (e.g., if "PFOA" is a node derived from \textit{chemicals analyzed}, values like "impaired kidney function" from a \textit{health outcome} dimension in the same paper can be associated with the paper node linked to "PFOA"). Importantly, this enriched structure also allows a researcher starting from a specific value in the \textit{health outcome} dimension (e.g., "impaired kidney function") to trace back and identify which entries from the \textit{chemicals analyzed} dimension (from the same or related documents) are associated with it, effectively highlighting potential causal or correlational factors. This bidirectional exploration, mirroring human cognition in linking multifaceted attributes to core concepts, provides a holistic knowledge view.

\section{Discussion}
\label{sec:discussion}

The KnoVo system demonstrate a important step towards nuanced, automated research novelty assessment. This discussion addresses its computational performance, key methodological considerations arising from its design and data handling, and its broader context and future applicability.

KnoVo's current implementation, while leveraging accessible local LLMs, involves computational costs detailed in Table~\ref{tab:component_timings}, particularly for deep 2-layer network analyses. The processing time is largely affected by the number of LLM interactions required, with key operations like value extraction ($\Lambda_{\text{extract}}$) and comparative analysis ($\Lambda_{\text{compare}}$) being invoked for numerous papers within a network. To enhance scalability, a multi-pronged optimization strategy is envisioned. This includes algorithmic improvements (e.g., sophisticated asynchronous processing, parallelization of independent tasks, LLM call batching, robust caching), model-level efficiencies (exploring smaller, faster, or quantized LLMs), and, most transformatively, the development of a pre-computed knowledge graph backend. Such a backend would minimize real-time LLM calls, shifting KnoVo towards reliance on efficient graph queries for many operations.

Beyond computational aspects, several methodological design choices and data considerations are pertinent to KnoVo's operation. Its analytical depth is currently based on abstracts; consequently, the absence of abstracts for some papers in scholarly databases like Semantic Scholar can limit the scope of analysis for those specific nodes. A core ability of KnoVo is its dynamic dimension extraction via an LLM function ($\Lambda_{\text{extract}}$, Sec \ref{subsec:dimension_extraction}). Crucially, to ensure relevance and expert alignment, KnoVo supports a \texttt{human-in-the-loop approach}, allowing users to add, remove, or refine these LLM-generated dimensions, thereby tailoring the analysis to specific research questions and validating the framework's inputs. Furthermore, practical restrictions in data acquisition, such as API rate limits from sources like Semantic Scholar, necessitate capping the number of processed citations. While this may establish selection bias, we mitigate this by employing relevance-based sorting for citation retrieval (as detailed in Sec~\ref{sec:dataset}), which promotes a more temporally and contextually diverse set of citing papers compared to purely chronological sorting.

In the broader context of bibliometric tools, it is important to admit that KnoVo operates within an ecosystem where citation networks can, in principle, be manipulated—an external factor that content-focused analysis like KnoVo's aims to look beyond by assessing intrinsic novelty. While this study focuses on formal citation networks, KnoVo’s core approach—dynamic dimension extraction and comparative analysis—is broadly applicable. It can effectively extend to semantically related document sets found through advanced search, enabling deeper content analysis beyond traditional citation structures.

\section{Future Work}

\noindent The current KnoVo system provides a robust foundation for automated novelty assessment, and the promising abilities of Large Language Models in scholarly analysis encourage several directions for its further development and refinement.

\textbf{Enhanced Content Scope.}
KnoVo currently relies on abstracts, which, while concise, may not capture a paper's full contribution. Future work will focus on incorporating full-text analysis. This will enable a more extensive assessment, though it requires tackling challenges of increased computational cost, efficient information retrieval from longer texts, and maintaining analytical focus.

\textbf{Advanced LLM Methodologies.}
KnoVo's performance is essentially linked to the underlying LLMs. We aim to examine a range of LLM architectures, data sources, and prompting methods. Particular attention will be given to combining models or prompts through ensemble or voting techniques to strengthen dimension extraction and comparison, while minimizing the influence of individual model bias.

\textbf{An Interactive Knowledge Navigation Platform.}
We envision KnoVo evolving into an interactive platform that allows users to explore pre-computed knowledge graphs or upload their own datasets, dynamically modifying visualizations and parameters to smoothly traverse the evolution of concepts. Such a system promises to greatly increase the effectiveness, originality, and impact of research and investigation across all domains by making the complex structure of scientific knowledge accessible and queryable. Thus, KnoVo provides a strong, data-driven framework for comprehending and expanding scientific understanding.

\section{Conclusion}
\label{sec:conclusion}

This paper introduced KnoVo, an intelligent system for the automated, quantitative assessment and analysis of research novelty within scientific literature. KnoVo goes beyond conventional impact metrics and subjective reviews to address the difficulty of navigating the growing corpus of academic literature. KnoVo provides a multi-dimensional and temporally-aware analysis of a paper's novelty in relation to both previous and subsequent research by utilizing Large Language Models (LLMs) to dynamically extract fine-grained dimensions of contribution and carry out nuanced, context-aware comparisons across multi-layered citation networks. Through the use of their 2-layer citation networks, we empirically demonstrated KnoVo's ability to quantify novelty, track the evolution of research ideas along specific dimensions, identify works with similar novelty profiles, highlight important advancements, and enable deeper exploration of complex research landscapes on a diverse dataset of twenty target papers from various scientific fields. KnoVo's accessibility-focused design, which makes use of local, open-source LLMs, is an important component.

This work offers several contributions: a novel conceptual framework for multi-dimensional novelty assessment; a functional prototype system (KnoVo); methodological advancements in dynamic dimension extraction from abstracts, LLM-driven comparative scoring, and temporal novelty tracking; an empirical validation of KnoVo's capabilities; and a comparative evaluation of relevant open-source LLMs for these tasks. Future studies will concentrate on developing LLM approaches with ensemble techniques and adaptive scoring mechanisms, expanding KnoVo's analytical depth through full-text analysis, carrying out thorough longitudinal and task-oriented validation, and eventually implementing KnoVo as an interactive knowledge navigation platform based on a scalable knowledge graph backend. KnoVo represents a promising, data-driven paradigm for enriching our understanding of the scientific landscape and accelerating knowledge discovery.



\section{Acknowledgment}

This Research was supported in part by a National Institutes of Health IDeA grant P20GM103408, a National Science Foundation CSSI grant OAC 2410668, and a US Department of Energy grant DE-0011014.




\printcredits

 \bibliographystyle{model1-num-names}


\bibliography{KnoVo-references}

\begin{thebibliography}{59}
\expandafter\ifx\csname natexlab\endcsname\relax\def\natexlab#1{#1}\fi
\providecommand{\url}[1]{\texttt{#1}}
\providecommand{\href}[2]{#2}
\providecommand{\path}[1]{#1}
\providecommand{\DOIprefix}{doi:}
\providecommand{\ArXivprefix}{arXiv:}
\providecommand{\URLprefix}{URL: }
\providecommand{\Pubmedprefix}{pmid:}
\providecommand{\doi}[1]{\href{http://dx.doi.org/#1}{\path{#1}}}
\providecommand{\Pubmed}[1]{\href{pmid:#1}{\path{#1}}}
\providecommand{\bibinfo}[2]{#2}
\ifx\xfnm\relax \def\xfnm[#1]{\unskip,\space#1}\fi
\bibitem[{Zhao and Zhang(2025)}]{zhao2025review}
\bibinfo{author}{Y.~Zhao}, \bibinfo{author}{C.~Zhang},
\newblock \bibinfo{title}{A review on the novelty measurements of academic papers},
\newblock \bibinfo{journal}{Scientometrics}  (\bibinfo{year}{2025}) \bibinfo{pages}{1--27}.
\bibitem[{Amplayo et~al.(2019)Amplayo, Hwang, and Song}]{amplayo2019evaluating}
\bibinfo{author}{R.~K. Amplayo}, \bibinfo{author}{S.-w. Hwang}, \bibinfo{author}{M.~Song},
\newblock \bibinfo{title}{Evaluating research novelty detection: Counterfactual approaches},
\newblock in: \bibinfo{booktitle}{Proceedings of the thirteenth workshop on graph-based methods for natural language processing (TextGraphs-13)}, \bibinfo{year}{2019}, pp. \bibinfo{pages}{124--133}.
\bibitem[{Yan et~al.(2020)Yan, Tian, and Zhang}]{YanTZ20}
\bibinfo{author}{Y.~Yan}, \bibinfo{author}{S.~Tian}, \bibinfo{author}{J.~Zhang},
\newblock \bibinfo{title}{The impact of a paper's new combinations and new components on its citation},
\newblock \bibinfo{journal}{Scientometrics} \bibinfo{volume}{122} (\bibinfo{year}{2020}) \bibinfo{pages}{895--913}.
\bibitem[{Foster et~al.(2021)Foster, Shi, and Evans}]{foster2021surprise}
\bibinfo{author}{J.~G. Foster}, \bibinfo{author}{F.~Shi}, \bibinfo{author}{J.~Evans},
\newblock \bibinfo{title}{Surprise! measuring novelty as expectation violation}  (\bibinfo{year}{2021}).
\bibitem[{Hou et~al.(2022)Hou, Wang, and Li}]{hou2022new}
\bibinfo{author}{J.~Hou}, \bibinfo{author}{D.~Wang}, \bibinfo{author}{J.~Li},
\newblock \bibinfo{title}{A new method for measuring the originality of academic articles based on knowledge units in semantic networks},
\newblock \bibinfo{journal}{Journal of Informetrics} \bibinfo{volume}{16} (\bibinfo{year}{2022}) \bibinfo{pages}{101306}.
\bibitem[{Cohen(2017)}]{cohen2017should}
\bibinfo{author}{B.~A. Cohen},
\newblock \bibinfo{title}{How should novelty be valued in science?},
\newblock \bibinfo{journal}{Elife} \bibinfo{volume}{6} (\bibinfo{year}{2017}) \bibinfo{pages}{e28699}.
\bibitem[{Ivancovsky et~al.(2024)Ivancovsky, Baror, and Bar}]{Ivancovsky_Baror_Bar_2024}
\bibinfo{author}{T.~Ivancovsky}, \bibinfo{author}{S.~Baror}, \bibinfo{author}{M.~Bar},
\newblock \bibinfo{title}{A shared novelty-seeking basis for creativity and curiosity},
\newblock \bibinfo{journal}{Behavioral and Brain Sciences} \bibinfo{volume}{47} (\bibinfo{year}{2024}) \bibinfo{pages}{e89}.
\bibitem[{Thelwall and Sud(2022)}]{thelwall2022scopus}
\bibinfo{author}{M.~Thelwall}, \bibinfo{author}{P.~Sud},
\newblock \bibinfo{title}{Scopus 1900--2020: Growth in articles, abstracts, countries, fields, and journals},
\newblock \bibinfo{journal}{Quantitative Science Studies} \bibinfo{volume}{3} (\bibinfo{year}{2022}) \bibinfo{pages}{37--50}.
\bibitem[{Vaswani et~al.(2017)Vaswani, Shazeer, Parmar, Uszkoreit, Jones, Gomez, Kaiser, and Polosukhin}]{vaswani2017attention}
\bibinfo{author}{A.~Vaswani}, \bibinfo{author}{N.~Shazeer}, \bibinfo{author}{N.~Parmar}, \bibinfo{author}{J.~Uszkoreit}, \bibinfo{author}{L.~Jones}, \bibinfo{author}{A.~N. Gomez}, \bibinfo{author}{{\L}.~Kaiser}, \bibinfo{author}{I.~Polosukhin},
\newblock \bibinfo{title}{Attention is all you need},
\newblock \bibinfo{journal}{Advances in neural information processing systems} \bibinfo{volume}{30} (\bibinfo{year}{2017}).
\bibitem[{Wei et~al.(2019)Wei, Hu, and Xing}]{10.1007/978-3-030-29551-6_58}
\bibinfo{author}{X.~Wei}, \bibinfo{author}{Y.~Hu}, \bibinfo{author}{L.~Xing},
\newblock \bibinfo{title}{Gated self-attentive encoder for neural machine translation},
\newblock in: \bibinfo{booktitle}{Knowledge Science, Engineering and Management: 12th International Conference, KSEM 2019, Athens, Greece, August 28–30, 2019, Proceedings, Part I}, \bibinfo{publisher}{Springer-Verlag}, \bibinfo{address}{Berlin, Heidelberg}, \bibinfo{year}{2019}, p. \bibinfo{pages}{655–666}. \URLprefix \url{https://doi.org/10.1007/978-3-030-29551-6_58}. \DOIprefix\doi{10.1007/978-3-030-29551-6_58}.
\bibitem[{Funk and Owen-Smith(2017)}]{funk2017dynamic}
\bibinfo{author}{R.~J. Funk}, \bibinfo{author}{J.~Owen-Smith},
\newblock \bibinfo{title}{A dynamic network measure of technological change},
\newblock \bibinfo{journal}{Management science} \bibinfo{volume}{63} (\bibinfo{year}{2017}) \bibinfo{pages}{791--817}.
\bibitem[{Bu et~al.(2021)Bu, Waltman, and Huang}]{bu2021multidimensional}
\bibinfo{author}{Y.~Bu}, \bibinfo{author}{L.~Waltman}, \bibinfo{author}{Y.~Huang},
\newblock \bibinfo{title}{A multidimensional framework for characterizing the citation impact of scientific publications},
\newblock \bibinfo{journal}{Quantitative science studies} \bibinfo{volume}{2} (\bibinfo{year}{2021}) \bibinfo{pages}{155--183}.
\bibitem[{Hofstra et~al.(2020)Hofstra, Kulkarni, Munoz-Najar~Galvez, He, Jurafsky, and McFarland}]{hofstra2020diversity}
\bibinfo{author}{B.~Hofstra}, \bibinfo{author}{V.~V. Kulkarni}, \bibinfo{author}{S.~Munoz-Najar~Galvez}, \bibinfo{author}{B.~He}, \bibinfo{author}{D.~Jurafsky}, \bibinfo{author}{D.~A. McFarland},
\newblock \bibinfo{title}{The diversity--innovation paradox in science},
\newblock \bibinfo{journal}{Proceedings of the National Academy of Sciences} \bibinfo{volume}{117} (\bibinfo{year}{2020}) \bibinfo{pages}{9284--9291}.
\bibitem[{Guo et~al.(2025)Guo, Yang, Zhang, Song, Zhang, Xu, Zhu, Ma, Wang, Bi et~al.}]{guo2025deepseek}
\bibinfo{author}{D.~Guo}, \bibinfo{author}{D.~Yang}, \bibinfo{author}{H.~Zhang}, \bibinfo{author}{J.~Song}, \bibinfo{author}{R.~Zhang}, \bibinfo{author}{R.~Xu}, \bibinfo{author}{Q.~Zhu}, \bibinfo{author}{S.~Ma}, \bibinfo{author}{P.~Wang}, \bibinfo{author}{X.~Bi}, et~al.,
\newblock \bibinfo{title}{Deepseek-r1: Incentivizing reasoning capability in llms via reinforcement learning},
\newblock \bibinfo{journal}{arXiv preprint arXiv:2501.12948}  (\bibinfo{year}{2025}).
\bibitem[{Team(2025)}]{gemmateam2025gemma3technicalreport}
\bibinfo{author}{G.~Team}, \bibinfo{title}{Gemma 3 technical report}, \bibinfo{year}{2025}. \URLprefix \url{https://arxiv.org/abs/2503.19786}. \href{http://arxiv.org/abs/2503.19786}{\tt arXiv:2503.19786}.
\bibitem[{Dubey et~al.(2024)Dubey, Jauhri, Pandey, Kadian, Al-Dahle, Letman, Mathur, Schelten, Yang, Fan et~al.}]{dubey2024llama}
\bibinfo{author}{A.~Dubey}, \bibinfo{author}{A.~Jauhri}, \bibinfo{author}{A.~Pandey}, \bibinfo{author}{A.~Kadian}, \bibinfo{author}{A.~Al-Dahle}, \bibinfo{author}{A.~Letman}, \bibinfo{author}{A.~Mathur}, \bibinfo{author}{A.~Schelten}, \bibinfo{author}{A.~Yang}, \bibinfo{author}{A.~Fan}, et~al.,
\newblock \bibinfo{title}{The llama 3 herd of models},
\newblock \bibinfo{journal}{arXiv preprint arXiv:2407.21783}  (\bibinfo{year}{2024}).
\bibitem[{{Mistral AI}(2025)}]{mistral2025small31}
\bibinfo{author}{{Mistral AI}}, \bibinfo{title}{{Mistral Small 3.1}}, \bibinfo{year}{2025}. \URLprefix \url{https://mistral.ai/news/mistral-small-3-1}, \bibinfo{note}{accessed: 2025-04-30}.
\bibitem[{Foster et~al.(2015)Foster, Rzhetsky, and Evans}]{foster2015tradition}
\bibinfo{author}{J.~G. Foster}, \bibinfo{author}{A.~Rzhetsky}, \bibinfo{author}{J.~A. Evans},
\newblock \bibinfo{title}{Tradition and innovation in scientists’ research strategies},
\newblock \bibinfo{journal}{American sociological review} \bibinfo{volume}{80} (\bibinfo{year}{2015}) \bibinfo{pages}{875--908}.
\bibitem[{Wagner et~al.(2019)Wagner, Whetsell, and Mukherjee}]{WAGNER20191260}
\bibinfo{author}{C.~S. Wagner}, \bibinfo{author}{T.~A. Whetsell}, \bibinfo{author}{S.~Mukherjee},
\newblock \bibinfo{title}{International research collaboration: Novelty, conventionality, and atypicality in knowledge recombination},
\newblock \bibinfo{journal}{Research Policy} \bibinfo{volume}{48} (\bibinfo{year}{2019}) \bibinfo{pages}{1260--1270}.
\bibitem[{Shibayama et~al.(2021)Shibayama, Yin, and Matsumoto}]{shibayama2021measuring}
\bibinfo{author}{S.~Shibayama}, \bibinfo{author}{D.~Yin}, \bibinfo{author}{K.~Matsumoto},
\newblock \bibinfo{title}{Measuring novelty in science with word embedding},
\newblock \bibinfo{journal}{PloS one} \bibinfo{volume}{16} (\bibinfo{year}{2021}) \bibinfo{pages}{e0254034}.
\bibitem[{Lin et~al.(2024)Lin, Peng, and Fang}]{lin2024evaluatingenhancinglargelanguage}
\bibinfo{author}{E.~Lin}, \bibinfo{author}{Z.~Peng}, \bibinfo{author}{Y.~Fang}, \bibinfo{title}{Evaluating and enhancing large language models for novelty assessment in scholarly publications}, \bibinfo{year}{2024}. \URLprefix \url{https://arxiv.org/abs/2409.16605}. \href{http://arxiv.org/abs/2409.16605}{\tt arXiv:2409.16605}.
\bibitem[{OpenAI et~al.(2024)OpenAI, Achiam, Adler, and Others}]{openai2024report}
\bibinfo{author}{OpenAI}, \bibinfo{author}{J.~Achiam}, \bibinfo{author}{S.~Adler}, \bibinfo{author}{Others}, \bibinfo{title}{Gpt-4 technical report}, \bibinfo{year}{2024}. \URLprefix \url{https://arxiv.org/abs/2303.08774}. \href{http://arxiv.org/abs/2303.08774}{\tt arXiv:2303.08774}.
\bibitem[{Amplayo et~al.(2018)Amplayo, Hong, and Song}]{AMPLAYO2018542}
\bibinfo{author}{R.~K. Amplayo}, \bibinfo{author}{S.~Hong}, \bibinfo{author}{M.~Song},
\newblock \bibinfo{title}{Network-based approach to detect novelty of scholarly literature},
\newblock \bibinfo{journal}{Information Sciences} \bibinfo{volume}{422} (\bibinfo{year}{2018}) \bibinfo{pages}{542--557}.
\bibitem[{{Martín de Diego} et~al.(2021){Martín de Diego}, González-Fernández, Fernández-Isabel, Fernández, and Cabezas}]{MARTINDEDIEGO2021101188}
\bibinfo{author}{I.~{Martín de Diego}}, \bibinfo{author}{C.~González-Fernández}, \bibinfo{author}{A.~Fernández-Isabel}, \bibinfo{author}{R.~R. Fernández}, \bibinfo{author}{J.~Cabezas},
\newblock \bibinfo{title}{System for evaluating the reliability and novelty of medical scientific papers},
\newblock \bibinfo{journal}{Journal of Informetrics} \bibinfo{volume}{15} (\bibinfo{year}{2021}) \bibinfo{pages}{101188}.
\bibitem[{Uzzi et~al.(2013)Uzzi, Mukherjee, Stringer, and Jones}]{uzzi2013atypical}
\bibinfo{author}{B.~Uzzi}, \bibinfo{author}{S.~Mukherjee}, \bibinfo{author}{M.~Stringer}, \bibinfo{author}{B.~Jones},
\newblock \bibinfo{title}{Atypical combinations and scientific impact},
\newblock \bibinfo{journal}{Science} \bibinfo{volume}{342} (\bibinfo{year}{2013}) \bibinfo{pages}{468--472}.
\bibitem[{Wang et~al.(2017)Wang, Veugelers, and Stephan}]{WANG20171416}
\bibinfo{author}{J.~Wang}, \bibinfo{author}{R.~Veugelers}, \bibinfo{author}{P.~Stephan},
\newblock \bibinfo{title}{Bias against novelty in science: A cautionary tale for users of bibliometric indicators},
\newblock \bibinfo{journal}{Research Policy} \bibinfo{volume}{46} (\bibinfo{year}{2017}) \bibinfo{pages}{1416--1436}.
\bibitem[{Chen and Fang(2019)}]{chen2019automatic}
\bibinfo{author}{L.~Chen}, \bibinfo{author}{H.~Fang},
\newblock \bibinfo{title}{An automatic method for extracting innovative ideas based on the scopus{\textregistered} database},
\newblock \bibinfo{journal}{KO KNOWLEDGE ORGANIZATION} \bibinfo{volume}{46} (\bibinfo{year}{2019}) \bibinfo{pages}{171--186}.
\bibitem[{Jeon et~al.(2023)Jeon, Lee, Ahn, and Lee}]{JEON2023101450}
\bibinfo{author}{D.~Jeon}, \bibinfo{author}{J.~Lee}, \bibinfo{author}{J.~M. Ahn}, \bibinfo{author}{C.~Lee},
\newblock \bibinfo{title}{Measuring the novelty of scientific publications: A fasttext and local outlier factor approach},
\newblock \bibinfo{journal}{Journal of Informetrics} \bibinfo{volume}{17} (\bibinfo{year}{2023}) \bibinfo{pages}{101450}.
\bibitem[{Wang et~al.(2024)Wang, Zhang, Chen, and Chen}]{Wang2023Novelty}
\bibinfo{author}{Z.~Wang}, \bibinfo{author}{H.~Zhang}, \bibinfo{author}{J.~Chen}, \bibinfo{author}{H.~Chen},
\newblock \bibinfo{title}{Measuring the novelty of scientific literature through contribution sentence analysis using deep learning and cloud model},
\newblock \bibinfo{journal}{SSRN Electronic Journal}  (\bibinfo{year}{2024}).
\bibitem[{Hinton(1986)}]{hinton1986learning}
\bibinfo{author}{G.~E. Hinton},
\newblock \bibinfo{title}{Learning distributed representations of concepts},
\newblock in: \bibinfo{booktitle}{Proceedings of the Annual Meeting of the Cognitive Science Society}, volume~\bibinfo{volume}{8}, \bibinfo{year}{1986}.
\bibitem[{Durbin et~al.(1998)Durbin, Eddy, Krogh, and Mitchison}]{Durbin_Eddy_Krogh_Mitchison_1998}
\bibinfo{author}{R.~Durbin}, \bibinfo{author}{S.~R. Eddy}, \bibinfo{author}{A.~Krogh}, \bibinfo{author}{G.~Mitchison}, \bibinfo{title}{Biological Sequence Analysis: Probabilistic Models of Proteins and Nucleic Acids}, \bibinfo{publisher}{Cambridge University Press}, \bibinfo{year}{1998}.
\bibitem[{OpenAI(2023)}]{openai_function_calling}
\bibinfo{author}{OpenAI}, \bibinfo{title}{Function calling in the openai api}, \bibinfo{howpublished}{\url{https://platform.openai.com/docs/guides/function-calling}}, \bibinfo{year}{2023}. \bibinfo{note}{Accessed: 2025-06-14}.
\bibitem[{Ammar et~al.(2018)Ammar, Groeneveld, Bhagavatula, Beltagy, Crawford, Downey, Dunkelberger, Elgohary, Feldman, Ha, Kinney, Kohlmeier, Lo, Murray, Ooi, Peters, Power, Skjonsberg, Wang, Wilhelm, Yuan, van Zuylen, and Etzioni}]{ammar-etal-2018-construction}
\bibinfo{author}{W.~Ammar}, \bibinfo{author}{D.~Groeneveld}, \bibinfo{author}{C.~Bhagavatula}, \bibinfo{author}{I.~Beltagy}, \bibinfo{author}{M.~Crawford}, \bibinfo{author}{D.~Downey}, \bibinfo{author}{J.~Dunkelberger}, \bibinfo{author}{A.~Elgohary}, \bibinfo{author}{S.~Feldman}, \bibinfo{author}{V.~Ha}, \bibinfo{author}{R.~Kinney}, \bibinfo{author}{S.~Kohlmeier}, \bibinfo{author}{K.~Lo}, \bibinfo{author}{T.~Murray}, \bibinfo{author}{H.-H. Ooi}, \bibinfo{author}{M.~Peters}, \bibinfo{author}{J.~Power}, \bibinfo{author}{S.~Skjonsberg}, \bibinfo{author}{L.~L. Wang}, \bibinfo{author}{C.~Wilhelm}, \bibinfo{author}{Z.~Yuan}, \bibinfo{author}{M.~van Zuylen}, \bibinfo{author}{O.~Etzioni},
\newblock \bibinfo{title}{Construction of the literature graph in semantic scholar},
\newblock in: \bibinfo{editor}{S.~Bangalore}, \bibinfo{editor}{J.~Chu-Carroll}, \bibinfo{editor}{Y.~Li} (Eds.), \bibinfo{booktitle}{Proceedings of the 2018 Conference of the North {A}merican Chapter of the Association for Computational Linguistics: Human Language Technologies, Volume 3 (Industry Papers)}, \bibinfo{publisher}{Association for Computational Linguistics}, \bibinfo{address}{New Orleans - Louisiana}, \bibinfo{year}{2018}, pp. \bibinfo{pages}{84--91}. \URLprefix \url{https://aclanthology.org/N18-3011/}. \DOIprefix\doi{10.18653/v1/N18-3011}.
\bibitem[{Berners{-}Lee(2011)}]{DBLP:conf/www/Berners-Lee11}
\bibinfo{author}{T.~Berners{-}Lee},
\newblock \bibinfo{title}{Designing the web for an open society},
\newblock in: \bibinfo{booktitle}{Proceedings of the 20th International Conference on World Wide Web, {WWW} 2011, Hyderabad, India, March 28 - April 1, 2011}, \bibinfo{year}{2011}, pp. \bibinfo{pages}{3--4}. \URLprefix \url{http://doi.acm.org/10.1145/1963405.1963408}. \DOIprefix\doi{10.1145/1963405.1963408}.
\bibitem[{{Google Scholar}(2025)}]{googlescholar}
\bibinfo{author}{{Google Scholar}}, \bibinfo{title}{Google scholar}, \bibinfo{howpublished}{\url{https://scholar.google.com/}}, \bibinfo{year}{2025}. \bibinfo{note}{Accessed: 2025-05-19}.
\bibitem[{OpenAI(2024)}]{openai2024gpt4technicalreport}
\bibinfo{author}{OpenAI}, \bibinfo{title}{Gpt-4 technical report}, \bibinfo{year}{2024}. \URLprefix \url{https://arxiv.org/abs/2303.08774}. \href{http://arxiv.org/abs/2303.08774}{\tt arXiv:2303.08774}.
\bibitem[{Anthropic(2024)}]{anthropic2024claude3}
\bibinfo{author}{Anthropic},
\newblock \bibinfo{title}{The claude 3 model family: Opus, sonnet, haiku},
\newblock \bibinfo{journal}{Anthropic}  (\bibinfo{year}{2024}).
\bibitem[{Team et~al.(2024)Team, Riviere, Pathak, Sessa, Hardin, Bhupatiraju, Hussenot, Mesnard, Shahriari, Ram{\'e} et~al.}]{team2024gemma}
\bibinfo{author}{G.~Team}, \bibinfo{author}{M.~Riviere}, \bibinfo{author}{S.~Pathak}, \bibinfo{author}{P.~G. Sessa}, \bibinfo{author}{C.~Hardin}, \bibinfo{author}{S.~Bhupatiraju}, \bibinfo{author}{L.~Hussenot}, \bibinfo{author}{T.~Mesnard}, \bibinfo{author}{B.~Shahriari}, \bibinfo{author}{A.~Ram{\'e}}, et~al.,
\newblock \bibinfo{title}{Gemma 2: Improving open language models at a practical size},
\newblock \bibinfo{journal}{arXiv preprint arXiv:2408.00118}  (\bibinfo{year}{2024}).
\bibitem[{Churchill et~al.(2019)Churchill, Padon, Sharma, and Aiken}]{10.1145/3314221.3314596}
\bibinfo{author}{B.~Churchill}, \bibinfo{author}{O.~Padon}, \bibinfo{author}{R.~Sharma}, \bibinfo{author}{A.~Aiken},
\newblock \bibinfo{title}{Semantic program alignment for equivalence checking},
\newblock in: \bibinfo{booktitle}{Proceedings of the 40th ACM SIGPLAN Conference on Programming Language Design and Implementation}, PLDI 2019, \bibinfo{publisher}{Association for Computing Machinery}, \bibinfo{address}{New York, NY, USA}, \bibinfo{year}{2019}, p. \bibinfo{pages}{1027–1040}. \URLprefix \url{https://doi.org/10.1145/3314221.3314596}. \DOIprefix\doi{10.1145/3314221.3314596}.
\bibitem[{Vaswani et~al.(2023)Vaswani, Shazeer, Parmar, Uszkoreit, Jones, Gomez, Kaiser, and Polosukhin}]{vaswani2023attentionneed}
\bibinfo{author}{A.~Vaswani}, \bibinfo{author}{N.~Shazeer}, \bibinfo{author}{N.~Parmar}, \bibinfo{author}{J.~Uszkoreit}, \bibinfo{author}{L.~Jones}, \bibinfo{author}{A.~N. Gomez}, \bibinfo{author}{L.~Kaiser}, \bibinfo{author}{I.~Polosukhin}, \bibinfo{title}{Attention is all you need}, \bibinfo{year}{2023}. \URLprefix \url{https://arxiv.org/abs/1706.03762}. \href{http://arxiv.org/abs/1706.03762}{\tt arXiv:1706.03762}.
\bibitem[{Devlin et~al.(2019)Devlin, Chang, Lee, and Toutanova}]{devlin2019bertpretrainingdeepbidirectional}
\bibinfo{author}{J.~Devlin}, \bibinfo{author}{M.-W. Chang}, \bibinfo{author}{K.~Lee}, \bibinfo{author}{K.~Toutanova}, \bibinfo{title}{Bert: Pre-training of deep bidirectional transformers for language understanding}, \bibinfo{year}{2019}. \URLprefix \url{https://arxiv.org/abs/1810.04805}. \href{http://arxiv.org/abs/1810.04805}{\tt arXiv:1810.04805}.
\bibitem[{Deng et~al.(2024)Deng, Zhang, Ding, Hu, Zhang, and Wang}]{deng2024zeroshotgeneralizableincrementallearning}
\bibinfo{author}{J.~Deng}, \bibinfo{author}{H.~Zhang}, \bibinfo{author}{K.~Ding}, \bibinfo{author}{J.~Hu}, \bibinfo{author}{X.~Zhang}, \bibinfo{author}{Y.~Wang}, \bibinfo{title}{Zero-shot generalizable incremental learning for vision-language object detection}, \bibinfo{year}{2024}. \URLprefix \url{https://arxiv.org/abs/2403.01680}. \href{http://arxiv.org/abs/2403.01680}{\tt arXiv:2403.01680}.
\bibitem[{DeCandia et~al.(2007)DeCandia, Hastorun, Jampani, Kakulapati, Lakshman, Pilchin, Sivasubramanian, Vosshall, and Vogels}]{10.1145/1294261.1294281}
\bibinfo{author}{G.~DeCandia}, \bibinfo{author}{D.~Hastorun}, \bibinfo{author}{M.~Jampani}, \bibinfo{author}{G.~Kakulapati}, \bibinfo{author}{A.~Lakshman}, \bibinfo{author}{A.~Pilchin}, \bibinfo{author}{S.~Sivasubramanian}, \bibinfo{author}{P.~Vosshall}, \bibinfo{author}{W.~Vogels},
\newblock \bibinfo{title}{Dynamo: amazon's highly available key-value store},
\newblock SOSP '07, \bibinfo{publisher}{Association for Computing Machinery}, \bibinfo{address}{New York, NY, USA}, \bibinfo{year}{2007}, p. \bibinfo{pages}{205–220}. \URLprefix \url{https://doi.org/10.1145/1294261.1294281}. \DOIprefix\doi{10.1145/1294261.1294281}.
\bibitem[{Dean and Ghemawat(2008)}]{10.1145/1327452.1327492}
\bibinfo{author}{J.~Dean}, \bibinfo{author}{S.~Ghemawat},
\newblock \bibinfo{title}{Mapreduce: simplified data processing on large clusters},
\newblock \bibinfo{journal}{Commun. ACM} \bibinfo{volume}{51} (\bibinfo{year}{2008}) \bibinfo{pages}{107–113}.
\bibitem[{Raasveldt and M\"{u}hleisen(2019)}]{10.1145/3299869.3320212}
\bibinfo{author}{M.~Raasveldt}, \bibinfo{author}{H.~M\"{u}hleisen},
\newblock \bibinfo{title}{Duckdb: an embeddable analytical database},
\newblock in: \bibinfo{booktitle}{Proceedings of the 2019 International Conference on Management of Data}, SIGMOD '19, \bibinfo{publisher}{Association for Computing Machinery}, \bibinfo{address}{New York, NY, USA}, \bibinfo{year}{2019}, p. \bibinfo{pages}{1981–1984}. \URLprefix \url{https://doi.org/10.1145/3299869.3320212}. \DOIprefix\doi{10.1145/3299869.3320212}.
\bibitem[{Lamb et~al.(2024)Lamb, Shen, Heres, Chakraborty, Kabak, Hsieh, and Sun}]{10.1145/3626246.3653368}
\bibinfo{author}{A.~Lamb}, \bibinfo{author}{Y.~Shen}, \bibinfo{author}{D.~Heres}, \bibinfo{author}{J.~Chakraborty}, \bibinfo{author}{M.~O. Kabak}, \bibinfo{author}{L.-C. Hsieh}, \bibinfo{author}{C.~Sun},
\newblock \bibinfo{title}{Apache arrow datafusion: A fast, embeddable, modular analytic query engine},
\newblock in: \bibinfo{booktitle}{Companion of the 2024 International Conference on Management of Data}, SIGMOD '24, \bibinfo{publisher}{Association for Computing Machinery}, \bibinfo{address}{New York, NY, USA}, \bibinfo{year}{2024}, p. \bibinfo{pages}{5–17}. \URLprefix \url{https://doi.org/10.1145/3626246.3653368}. \DOIprefix\doi{10.1145/3626246.3653368}.
\bibitem[{Lander et~al.(2001)Lander, Linton, Birren, Nusbaum, Zody, Baldwin, Devon, Dewar et~al.}]{Lander2001InitialSA}
\bibinfo{author}{E.~S. Lander}, \bibinfo{author}{L.~Linton}, \bibinfo{author}{B.~W. Birren}, \bibinfo{author}{C.~Nusbaum}, \bibinfo{author}{M.~C. Zody}, \bibinfo{author}{J.~Baldwin}, \bibinfo{author}{K.~Devon}, \bibinfo{author}{K.~Dewar}, et~al.,
\newblock \bibinfo{title}{Initial sequencing and analysis of the human genome.},
\newblock \bibinfo{journal}{Nature} \bibinfo{volume}{409 6822} (\bibinfo{year}{2001}) \bibinfo{pages}{860--921}.
\bibitem[{Jinek et~al.(2012)Jinek, Chylinski, Fonfara, Hauer, Doudna, and Charpentier}]{Jinek2012APD}
\bibinfo{author}{M.~Jinek}, \bibinfo{author}{K.~Chylinski}, \bibinfo{author}{I.~Fonfara}, \bibinfo{author}{M.~H. Hauer}, \bibinfo{author}{J.~A. Doudna}, \bibinfo{author}{E.~Charpentier},
\newblock \bibinfo{title}{A programmable dual-rna–guided dna endonuclease in adaptive bacterial immunity},
\newblock \bibinfo{journal}{Science} \bibinfo{volume}{337} (\bibinfo{year}{2012}) \bibinfo{pages}{816 -- 821}.
\bibitem[{Roosa et~al.(2020)Roosa, Lee, Luo, Kirpich, Rothenberg, Hyman, Yan, and Chowell}]{Roosa2020RealtimeFO}
\bibinfo{author}{K.~M. Roosa}, \bibinfo{author}{Y.~Lee}, \bibinfo{author}{R.~Luo}, \bibinfo{author}{A.~S. Kirpich}, \bibinfo{author}{R.~B. Rothenberg}, \bibinfo{author}{J.~M. Hyman}, \bibinfo{author}{P.~Yan}, \bibinfo{author}{G.~Chowell},
\newblock \bibinfo{title}{Real-time forecasts of the 2019-ncov epidemic in china from february 5th to february 24th, 2020},
\newblock \bibinfo{journal}{arXiv: Populations and Evolution}  (\bibinfo{year}{2020}).
\bibitem[{Christopher et~al.(2020)Christopher, Murray, and Rabiee}]{Christopher2020GlobalBO}
\bibinfo{author}{P.~Christopher}, \bibinfo{author}{J.~L. Murray}, \bibinfo{author}{N.~Rabiee},
\newblock \bibinfo{title}{Global burden of 369 diseases and injuries in 204 countries and territories, 1990–2019: a systematic analysis for the global burden of disease study 2019},
\newblock \bibinfo{journal}{Lancet (London, England)} \bibinfo{volume}{396} (\bibinfo{year}{2020}) \bibinfo{pages}{1204 -- 1222}.
\bibitem[{Kitaev(1997)}]{Kitaev1997FaultTQ}
\bibinfo{author}{A.~Y. Kitaev},
\newblock \bibinfo{title}{Fault tolerant quantum computation by anyons},
\newblock \bibinfo{journal}{Annals of Physics} \bibinfo{volume}{303} (\bibinfo{year}{1997}) \bibinfo{pages}{2--30}.
\bibitem[{Arute et~al.(2019)Arute, Arya, Babbush, Bacon, Bardin, Barends, Biswas, Boixo, Brand{\~a}o et~al.}]{Arute2019QuantumSU}
\bibinfo{author}{F.~Arute}, \bibinfo{author}{K.~Arya}, \bibinfo{author}{R.~Babbush}, \bibinfo{author}{D.~Bacon}, \bibinfo{author}{J.~C. Bardin}, \bibinfo{author}{R.~Barends}, \bibinfo{author}{R.~Biswas}, \bibinfo{author}{S.~Boixo}, \bibinfo{author}{F.~G. S.~L. Brand{\~a}o}, et~al.,
\newblock \bibinfo{title}{Quantum supremacy using a programmable superconducting processor},
\newblock \bibinfo{journal}{Nature} \bibinfo{volume}{574} (\bibinfo{year}{2019}) \bibinfo{pages}{505 -- 510}.
\bibitem[{Jordan et~al.(2011)Jordan, Lee, and Preskill}]{Jordan2011QuantumAF}
\bibinfo{author}{S.~P. Jordan}, \bibinfo{author}{K.~S.~M. Lee}, \bibinfo{author}{J.~Preskill},
\newblock \bibinfo{title}{Quantum algorithms for quantum field theories},
\newblock \bibinfo{journal}{Science} \bibinfo{volume}{336} (\bibinfo{year}{2011}) \bibinfo{pages}{1130 -- 1133}.
\bibitem[{Farrelly and Streich(2020)}]{Farrelly2020DiscretizingQF}
\bibinfo{author}{T.~Farrelly}, \bibinfo{author}{J.~Streich},
\newblock \bibinfo{title}{Discretizing quantum field theories for quantum simulation},
\newblock \bibinfo{journal}{arXiv: Quantum Physics}  (\bibinfo{year}{2020}).
\bibitem[{Rosenbaum and Rubin(1983)}]{Rosenbaum1983TheCR}
\bibinfo{author}{P.~R. Rosenbaum}, \bibinfo{author}{D.~B. Rubin},
\newblock \bibinfo{title}{The central role of the propensity score in observational studies for causal effects},
\newblock \bibinfo{journal}{Biometrika} \bibinfo{volume}{70} (\bibinfo{year}{1983}) \bibinfo{pages}{41--55}.
\bibitem[{Angrist and Pischke(2008)}]{Angrist2008MostlyHE}
\bibinfo{author}{J.~D. Angrist}, \bibinfo{author}{J.-S. Pischke},
\newblock \bibinfo{title}{Mostly harmless econometrics: An empiricist's companion},
\newblock \bibinfo{year}{2008}. \URLprefix \url{https://api.semanticscholar.org/CorpusID:63231051}.
\bibitem[{Mullainathan and Spiess(2017)}]{Mullainathan2017MachineLA}
\bibinfo{author}{S.~Mullainathan}, \bibinfo{author}{J.~Spiess},
\newblock \bibinfo{title}{Machine learning: An applied econometric approach},
\newblock \bibinfo{journal}{Journal of Economic Perspectives} \bibinfo{volume}{31} (\bibinfo{year}{2017}) \bibinfo{pages}{87--106}.
\bibitem[{Trudel et~al.(2008)Trudel, Horowitz, Wormuth, Scheringer, Cousins, and Hungerb{\"u}hler}]{Trudel2008EstimatingCE}
\bibinfo{author}{D.~Trudel}, \bibinfo{author}{L.~S. Horowitz}, \bibinfo{author}{M.~Wormuth}, \bibinfo{author}{M.~Scheringer}, \bibinfo{author}{I.~T. Cousins}, \bibinfo{author}{K.~Hungerb{\"u}hler},
\newblock \bibinfo{title}{Estimating consumer exposure to pfos and pfoa},
\newblock \bibinfo{journal}{Risk Analysis} \bibinfo{volume}{28} (\bibinfo{year}{2008}).
\bibitem[{Costa et~al.(2009)Costa, Sartori, and Consonni}]{Costa2009ThirtyYO}
\bibinfo{author}{G.~Costa}, \bibinfo{author}{S.~Sartori}, \bibinfo{author}{D.~Consonni},
\newblock \bibinfo{title}{Thirty years of medical surveillance in perfluooctanoic acid production workers},
\newblock \bibinfo{journal}{Journal of Occupational and Environmental Medicine} \bibinfo{volume}{51} (\bibinfo{year}{2009}) \bibinfo{pages}{364--372}.

\end{thebibliography}






\end{document}